\let\myo=\o
\documentclass[10pt]{article}
\usepackage{amssymb,amsmath,amsfonts,color,cite}
\usepackage{graphicx}
\usepackage{psfrag}
\usepackage[labelfont=bf,labelsep=period,justification=raggedright]{caption}
\usepackage[labelfont=bf,justification=raggedright,font=scriptsize,labelformat=empty,labelsep=colon]{subcaption}
\usepackage[section]{placeins}
\usepackage[finalnew]{trackchanges}

\makeatletter
\renewcommand{\@biblabel}[1]{\quad#1.}
\makeatother

\date{}

\definecolor{red}{rgb}{1,0,0}

\begin{document}

\begin{flushleft}
{\Large
\textbf{A computational model for histone mark propagation reproduces the distribution of heterochromatin in different human cell types}
}
% Insert Author names, affiliations and corresponding author email.
\\
Veit Schw\"ammle$^{\ast}$, 
Ole N\myo{}rregaard Jensen
\\
\bf Department for Biochemistry and Molecular Biology, University of Southern Denmark, Campusvej 55, DK-5230 Odense M,Denmark
\\
$\ast$ E-mail: veits@bmb.sdu.dk
\end{flushleft}
\section*{Abstract}
\noindent
Chromatin is a highly compact and dynamic nuclear structure that consists of DNA and associated proteins. The main organizational unit is the
nucleosome, which consists of a histone octamer with DNA wrapped around it.  Histone proteins are implicated in the regulation of eukaryote genes and
they carry numerous reversible post-translational modifications that control DNA-protein interactions and the recruitment of chromatin binding
proteins.  Heterochromatin, the transcriptionally inactive part of the genome, is densely packed and contains histone H3 that is methylated at Lys 9 
(H3K9me). The propagation of H3K9me in nucleosomes along the DNA in chromatin is antagonizing by methylation of H3 Lysine 4 (H3K4me) and acetylations
of several lysines, which is related to euchromatin and active genes. We show that the related histone modifications form antagonized domains on a
coarse scale. 
These histone marks are assumed to be initiated within distinct nucleation sites in the DNA and to propagate bi-directionally.
We propose a simple computer model that simulates the distribution of heterochromatin in human chromosomes. 
The simulations are in agreement with previously reported experimental observations from
two different human cell lines.
We reproduced different types of barriers between heterochromatin and euchromatin providing a unified model for their function. The effect of
changes in the nucleation site distribution and of propagation rates were studied. The former occurs mainly with the aim of
(de-)activation of single genes or gene groups and the latter has the power of controlling the transcriptional programs of entire
chromosomes. Generally, the regulatory program of gene transcription is controlled by the distribution of nucleation sites along the DNA string.

% \section*{Author Summary}
% Chromatin is a highly compact and dynamic nuclear structure that consists of DNA and associated proteins. The main organizational unit is the
% nucleosome, which consists of a histone octamer with DNA wrapped around it. 
%  Histone proteins are implicated in the regulation of eukaryote genes and 
%  they carry numerous reversible post-translational modifications that control DNA-protein interactions and the recruitment of chromatin binding
%  proteins. Chromatin can be classified into two main regimes consisting of domains along the chromosomes that are marked by respective
% histone modifications. Euchromatin denotes the transcriptionally active part, whereas genes are generally silenced in
% heterochromatin. We simulate the dynamics of the histone modification distribution with a stochastic model that builds on the processes nucleation,
% propagation and competition between the modifications within histones. The simulations match the data obtained by experiments
% for the histone modification distribution in human chromosomes. By extending the model to multiple coexisting modifications, we simulated different
% meta-stable cell states as well as oscillatory behavior.

\section*{Introduction}
Eukaryote DNA is organized in a highly compact structure, chromatin, that consists of deoxyribonucleic acids and proteins. The DNA double helix is
wound up around nucleosomes consisting of histone octamers, including two subunits each of histones H2A, H2B, H3 and H4. A plethora of
proteins are involved in maintaining and regulating chromatin structure during DNA replication, transcription, repair, etc. DNA methylation,
nucleosome positioning and reversible post-translational modifications of histone proteins govern the spatial
organization and accessibility of DNA in chromatin in eukaryote cells. The post-translational modifications of histones, also known as histone marks,
include methylation, acetylation, phosphorylation and other covalent chemical moieties that are (reversibly) conjugated to distinct amino acid
residues in the histone proteins. These site-specific and co-existing modifications of multiple amino acid residues generate complex combinatorial
patterns that may have functional roles in modulating chromatin structure and in the recruitment of specific protein co-factors to distinct domains in
chromatin, thereby constituting a highly dynamic regulatory network~\cite{Fodor2006}.
Heterochromatin denotes the highly condensed inactive state of chromatin, where genes are
repressed due to the inaccessibility of DNA for the transcription machinery. 
Abnormal function of the heterochromatic state has been linked to several diseases~\cite{Dialynas2008,Norwood2006,Cloos2006}.

In the present work we address several fundamental questions in chromatin biology and histone structure/function relationships:
(a) Are histone modifications organized in domains along the chromatin?
(b) What is the minimal model able to simulate the formation of heterochromatin domains that is in accordance with experimental results?
(c) What are the different mechanisms leading to changes of the histone modification landscape and which are able to switch genes on/off as response
to
external stimuli? 

Several computational and/or mathematical approaches simulate a bistable state of histone modifications, for example switching between a state
dominated by H3K9 methylation and the state dominated by H3K9 acetylations~\cite{Dodd2007a,Sneppen2008,Micheelsen2010,Sedighi2007}. These
studies concentrated on a general stability analysis and memory of such a system, thereby revealing
ultrasensitive switching behavior. However, there was no direct comparison of those results to experimentally measured chromatin configurations. In
another approach, the formation of multiply modified histones was described by stochastic nonlinear
equations~\cite{Gils2009}. The analysis did not consider specific modifications as the model only counted the number of modifications
on a histone without specifying their type. An epigenetic switch was modeled in ref.~\cite{Angel2011}, where the authors studied switching and memory
effects of the floral repressor of \emph{Arabidopsis} with a simple mathematical model implementing nucleation and spreading of the silencing
H3K27me3 mark. The data was successfully compared to ChIP data. Furthermore, simulations of the heterochromatin domain around the Oct4 locus in mouse
ES cells and fibroblasts showed that this domain and most euchromatic H3K9me3 domains were well-described by a model based on propagation of the
marks without taking into account specific boundary or insulator elements~\cite{Hathaway2012}.

We go further and simulate the formation of heterochromatin over whole human chromosomes. The computer model implements the basic
processes of nucleation, propagation and competition of histone marks through stochastic rates. We test whether such a
simple model is able to generate stable domains of competing histone modifications. We then compare the results to
experimental measurements and study the model's overall behavior.

In the following, we present biological evidence for the rules implemented in our computational model.

\noindent
{\bf Nucleation:}
Non-protein-coding DNA sequences seem to play a crucial role to nucleate histone modification mediated domain formation.
The RNA interference machinery
shows activity at dh-dg repeats in yeast DNA ~\cite{Hall2002,Grewal2007} leading to
heterochromatin formation through a self-amplifying feed-forward regulatory mechanism ~\cite{Sugiyama2005,Noma2004}.
In higher eukaryotes, details about the initialization of heterochromatin remain unclear but strong correlations between heterochromatin and
diverse satellite-repeats and transposable elements were
observed~\cite{Slotkin2007,Martens2005}, as for instance with SINE-Alu elements in humans~\cite{Kondo2003a}.
We will refer to these initiating sequences from now on as heterochromatin \emph{nucleation sites}~\cite{Cam2005}.
Within this scenario, these sequences contain regulatory information over gene transcription that can be fine-tuned to
allow the development of different cell types~\cite{Taft2007,Mattick2007}. We will show how this information can be used to generate
different heterochromatic states. 

The presence of genomic CpG islands is strongly correlated to transcriptional activity, and makes CpG islands candidates for nucleation sites for
transcriptionally activating histone marks. CpG islands exhibit a high abundance of demethylated DNA, enrichment of H3K4me2/3, H3K9ac
and H3K14ac marks ~\cite{Bock2007,Bernstein2006,Roh2005}. The underlying mechanisms involve the protein Cfp1 that associates with
unmethylated CpG islands \emph{in
vivo} and recruits H3K4 methyltransferases to nearby histones~\cite{Thomson2010}.
SINEAlu elements and CpG appear to be experimentally well characterized nucleation sites. Other types of nucleation sites were neglected in
the model.

According to these relations between DNA sequence and chromatin state, both repressive and activating histone marks are occurring
in specific gene regions and are initiated by their respective nucleation sites. However, different cell types exhibit
distinct transcriptome profiles reflected in different histone modifications~\cite{Heintzman2009}.
Within chromatin, an effective regulatory mechanism is required for switching transcriptional activity of large genomic regions and for the
formation
 of distinct transcription patterns. One might argue that the presence of predefined “fixed” nucleation sites for initiation of histone marks within
 the genome is incompatible with the presence of different transcriptional states. We demonstrate that predefined
 nucleation sites and histone modifications will indeed provide features that allow for a dynamic switching behavior of genes and genomic regions.

\noindent
{\bf Propagation:}
Chromosomes exhibit regions mediated by histone modifications that expand over considerable ranges along the DNA strand. 
Heterochromatin domains represent one type of these regions.
 Reinforcing mechanisms lead to the formation of heterochromatin \emph{domains} enriched in H3K9me2 and H3K9me3 marks \cite{Wit2009}.
Di- and trimethylation of lysine 9 on histone H3 by the methyltransferases G9a and Suv39h1 reflect the repressed state of
heterochromatin~\cite{Jenuwein2001,Peters2003,Rice2003} that is maintained by
several proteins through a positive feedback loop~\cite{Grewal2003,Maison2004}. Heterochromatin protein-1 (HP1) recognizes
H3K9 methylation~\cite{Schotta2002,James89,Lachner2001} and it interacts with Suv39h1~\cite{Czvitkovich2001,Aagaard1999}, that is
recruited by neighboring H3K9me2 sites~\cite{Lachner2001} and thereby stabilizes the
heterochromatin state. HP1 also recruits the DNA methyltransferases DNMT1 that itself associates with G9a. 
G9a sets the H3K9me2 mark~\cite{Esteve2006,Smallwood2007}.
HP1 establishes the spatially dense chromatin structure and recruits histone deacetylases and DNA methyltransferases to strengthen
this state~\cite{Fuks2003,Zhang2002}. This loop is further stimulated as HP1 associates to itself~\cite{Douarin1996} leading to the propagation of
H3K9me2 and H3K9me3.

Transcriptionally active regions contain not only H3K4me3 but also the other methylation states H3K4me2
and H3K4me1.  It was found that the USF 
protein binds to specific DNA sequences and mediates H3K4 methylation and histone acetylation~\cite{Gaszner2006}, which are both related to gene
activation. There is strong correlation between H3K4 methylations and e.g. the acetylation marks H3K14ac,
H3K18ac~\cite{Nightingale2007,Liu2005}. Generally, local recruitment of histone acetyltransferases activities seems to counteract the
spreading of heterochromatin~\cite{West2004,Chiu2003,Donze2002}. This whole machinery suggests a propagation mechanism for euchromatin formation.

\noindent
{\bf Competition:}
How are the efficient molecular processes for the propagation of heterochromatin domains
prevented from reaching a state where heterochromatin completely occupies the chromosomes? A total occupation would lead to a complete shut-down of
gene transcription. Boundary elements, also called insulators, are defined as genetic regions where the propagation of histone marks are stalled. In a
DNA strand with only nearest-neighbor interactions, the boundary must permanently flank both ends of the to-be-protected domain to shield it
from silencing~\cite{Pikaart1998}. 

Passive insulators that prevent the setting of H3K9 methylation or HP1 association without actively recruiting histone-modifying enzymes are
not able to stop heterochromatin propagation~\cite{Kimura2004}. 
The condensed three-dimensional conformation of the DNA strand allows propagation of heterochromatin marks between co-located non-neighboring
nucleosomes leading to the propagation into the to-be-insulated region. Furthermore, passive insulators would be required to form a static and
stable barrier. Otherwise, heterochromatin domains would be able to temporally
spread into the region. Hence, effective insulation can only work when the insulator
\emph{actively} maintains a region of histone marks that antagonize the setting of the heterochromatin forming constituents on the entire
region.

Wang et al.~\cite{Wang2012} present a model that explains both fixed boundaries, with a specific actively recruiting boundary element, and
broader ones where euchromatin related histone modifications gradually change in to the ones related to the repressive state. We will show that both
scenarios are possible in our simulations, mainly depending on the distribution of nucleation sites.

Histone marks that are related to gene activation seem also to be main players in controlling heterochromatin formation.
Domains decorated with H3 and H4 acetylation marks and the H3K4 methylation mark prevent heterochromatin from spreading over the entire
chromosomes~\cite{Mutskov2004,West2004}. 
H3K4me3 competes directly with the heterochromatin state as it inhibits the methylation of H3K9 by
Suv39h1~\cite{Zegerman2002,Nishioka2002}. Similarly, H3K9ac inhibits histone deacetylases and interrupts the interaction between HP1 and
chromatin~\cite{Taddei2001}.

After identifying the processes propagation, nucleation and competition as main actors in histone domain formation, it is possible to
construct a
theoretical model for the distribution of active and inactive chromatin domains by assuming that the competing histone domains become
initiated from their respective nucleation sites. We will test this assumption with a computational model that implements the basic underlying
rules.

\section*{Results}

\begin{figure}[!ht]
\includegraphics[angle=0 ,width=1.0\textwidth]{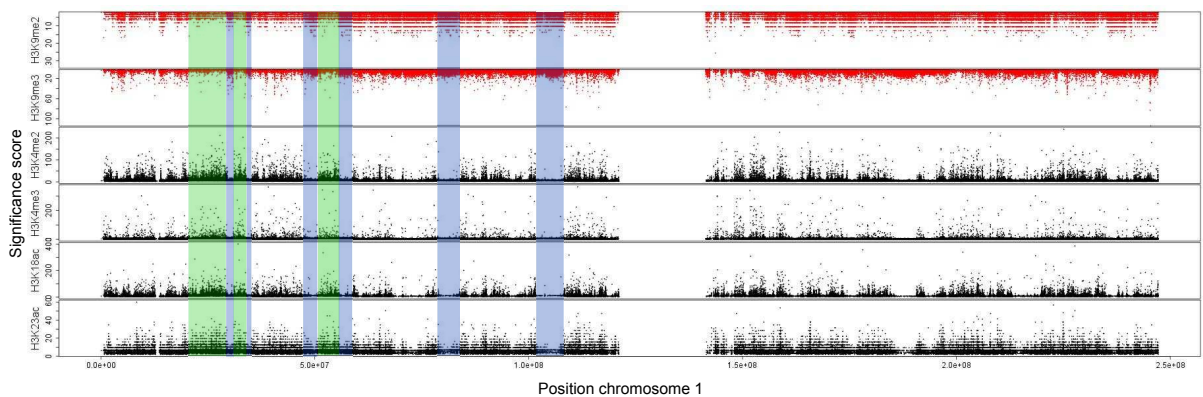}
\caption{{\bf Heterochromatic and euchromatic histone modifications form non-overlapping domains on a coarse scale.}
We used the measurements obtained in a genome-wide experiment on human CD4+ T cells from \cite{Barski2007a} and \cite{Wang2008}
and analyzed them using the CCAT (version 3.0) tool ~\cite{Xu2010}. 
We adapt the analysis using slightly less stringent parameters than the default ones allowing for noisy measurements with lower significance in
order to obtain a most complete modification landscape. 
The plot exhibits the significance scores of the histone modifications H3K9me2 and H3K9me3 related to heterochromatin (red) and H3K4me2, H3K4me3,
H3K18ac and H3K23ac related to euchromatin (black). We visualized the distribution over entire chromosome 1. Heterochromatin marks were plotted upside 
down for better visualization. Both sets of modifications form very similar patterns and form regions of higher and lower
abundance. We marked some of the regions with high (low) euchromatic and low (high) heterochromatic content green (blue). }
\label{fig:Bmeasured}
\end{figure}

\subsection*{Experimental evidence for the formation of chromatin domains}
ChIP-chip and ChIP-seq experiments are able to determine the genomic location of distinct histone marks~\cite{Park2009}. 
Measurements of the histone modification landscape of the human genome can lead to the identification of general patterns. 
Studies of human chromosomes showed a backbone of modifications at gene promoters~\cite{Wang2008} that exhibited different patterns for different
amounts of CpG islands in the promoters leading to a
rather confident prediction of gene expression by correlation to histone modification levels~\cite{Karlic2010}.

We focus here on patterns along entire chromosomes and investigate whether they remain independently of the biological function of the co-located DNA.
While it is known that the heterochromatic marks H3K9me2 and H3K9me3 form large chromatin domains, it is not clear whether the marks
corresponding to euchromatin also form domains on a coarse scale. 

We analyzed the ChIP-seq experiments of CD4+ T cells from refs.~\cite{Barski2007a,Wang2008} and HeLa cells from
refs.~\cite{Vermeulen2010,Liu2010,Qi2010}, applying the CCAT (version 3.0) tool on the downloaded raw data. The resulting scores give an estimate for
a histone modification to be at the corresponding genomic position. It can be clearly seen that the scores for the heterochromatin
marks H3K9me2 and H3K9me3 form landscapes that exhibit very similar shapes on a large scale (Figure~\ref{fig:Bmeasured} and supplementary
Figures~\ref{fig:Bmeasured2}-\ref{fig:Hmeasured17} in File S1). The same can be observed for the activating marks H3K4me2, H3K4me3, H3K18ac and H3K23ac.
Furthermore, it seems that regions with higher scores for the euchromatic marks come with lower ones for the heterochromatic marks and vice verse,
forming non-overlapping domains.

\begin{figure}[!ht]
\includegraphics[angle=0,width=0.9\textwidth]{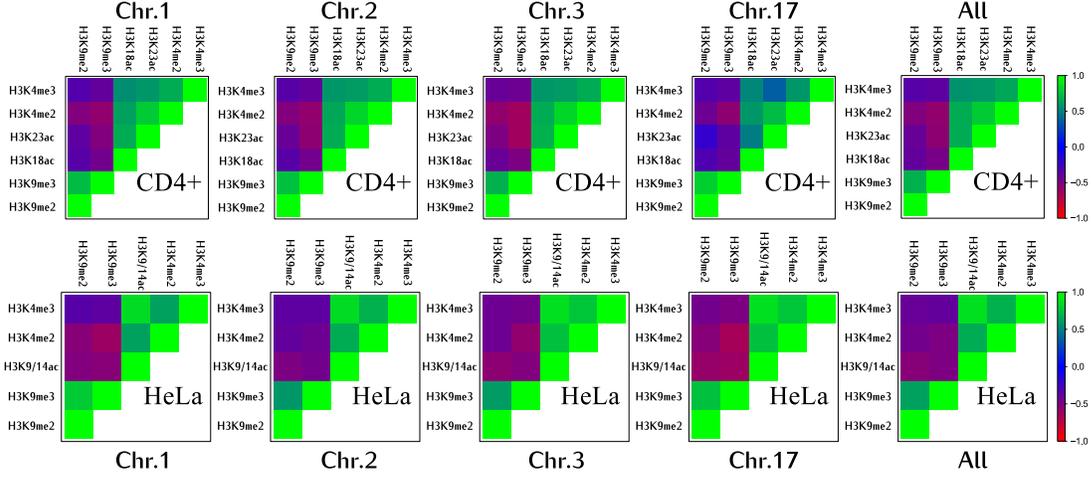}
\caption{{\bf Euchromatin and heterochromatin marks become anti-correlated on a coarse scale.} Pearson's correlation
between modifications within bins of 100 kbp for different chromosomes and two cell lines. Instead of taking into account the
individual scores in a bin, we simplify the content by the sum of all scores. It can be clearly observed that euchromatic (H3K9/14ac,
H3K18ac, H3K23ac, H3K4me2 and H3K4me3) and heterochromatic histone marks (H3K9me2, and H3K9me3) oppose each other for all considered chromosomes as
well as for two different cell lines. We therefore show that these modifications form long domains that are still detectable on a scale of about 1000
histones.}
\label{fig:measuredCorrs}
\end{figure}

In order to provide further evidence for this observation, we calculate Pearson's correlation between these marks on a coarse scale. Therefore, we
divide the chromosomes into bins of 100 kbp, taking for each bin the sum of the corresponding scores. The high (anti-)correlations between the
investigated histone modifications support the idea of large, non-overlapping domains for both heterochromatic and euchromatic histone modifications
(Figure~\ref{fig:measuredCorrs}). Hence, the size of these domains comes up to at least thousand histones (approximately 100 kbp).

\subsection*{Simulations of human chromosomes}

\begin{figure}[!ht]
\includegraphics[width=0.8\textwidth]{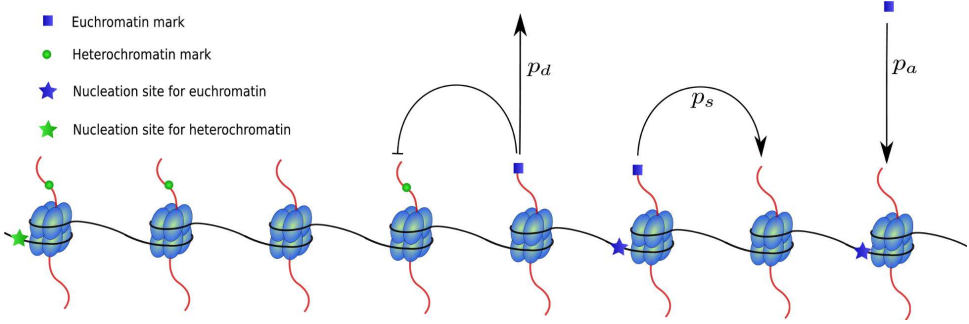}
\caption{{\bf Illustration of the processes nucleation, propagation, competition and deletion in the computational model.} Only nucleosomes
with nucleation sites can directly be modified with the respective modification with rate $p_a$ (probability per histone and time step). Empty
nucleosomes with neighboring modified nucleosomes obtain a modification of the same type with rate $p_s$. This parameter can vary for different
modification types (euchromatic or heterochromatic). Multiply modified nucleosomes are not allowed in the model with only competing marks and
therefore a new mark will not be set if the histone is already modified. Finally, every modified nucleosome looses its modification with rate $p_d$.}
\label{fig:scheme}
\end{figure}

The simulations are based on a minimal model for the formation of chromatin domains and will be directly compared to the ChIP-seq data studied in
the previous section. We converted the knowledge about the basic mechanisms of domain formation presented in the Introduction into computational
rules. 
The underlying stochastic processes of the computational approach are illustrated in
Figure~\ref{fig:scheme}. Dynamic setting and deleting of histone marks is based on the four processes nucleation, propagation, deletion and
competition: (i) histones can be directly modified at respective nucleation sites with rate $p_a$ (SINE-Alu elements for heterochromatin marks and CpG
islands for euchromatin marks); (ii) already modified histones propagate their modification to neighboring nucleosomes with rates $p_{s,1}$ and
$p_{s,2}$ for heterochromatin and euchromatin marks, respectively; (iii) histone modifications are removed with rate $p_d$; (iv) histones cannot be
simultaneously modified with heterochromatic and euchromatic marks. For details of the model, see section Methods. The model allows to simulate the
individual states of a large
number of nucleosomal units. The simulations do not take into account the effect of DNA replication involving the incorporation of new histone molecules across the chromosomes. The nucleosome states are updated each time step based on probabilistic rates, i.e. each
nucleosome changes its state between unmodified, or having either an euchromatin or a heterochromatin mark. The use of stochastic rates allows to generate
a highly dynamic state that can exhibit temporal fluctuations and long-range correlations. In this implementation of the model, there are no fixed
barriers that completely prevent propagation of a domain over it. Although we do not allow direct long-range interactions between
non-neighboring  nucleosomes, these nucleosomes may still interact through temporal fluctuations that might lead to e.g. collective changes of an
entire region.

We compare the results of the simulations for the distribution of human heterochromatin to experimental data from CD4+ T cells and HeLa cells.
By using the positioning data for the nucleation sites available in RepeatMasker~\cite{Smit1996} and the
UCSC Genome Browser~\cite{Fujita2011}, it is possible to include the
positions of SineAlu and CpG sites on the chromosomes as nucleation sites for heterochromatic and euchromatic marks, respectively. We
focus here on the distribution of the histone modifications on human chromosomes 1-3 and 17. Chromosome 17 was chosen to test how the model
performs for chromosomes with high GC-content. The distributions are collected from the simulations for the individual temporal changes of
the histone states of entire chromosomes. For instance, the simulations of the first human chromosome involve the simultaneous dynamics of about 2.5
million histones. 

\begin{figure}[ht!]
\includegraphics[angle=0,width=0.95\textwidth]{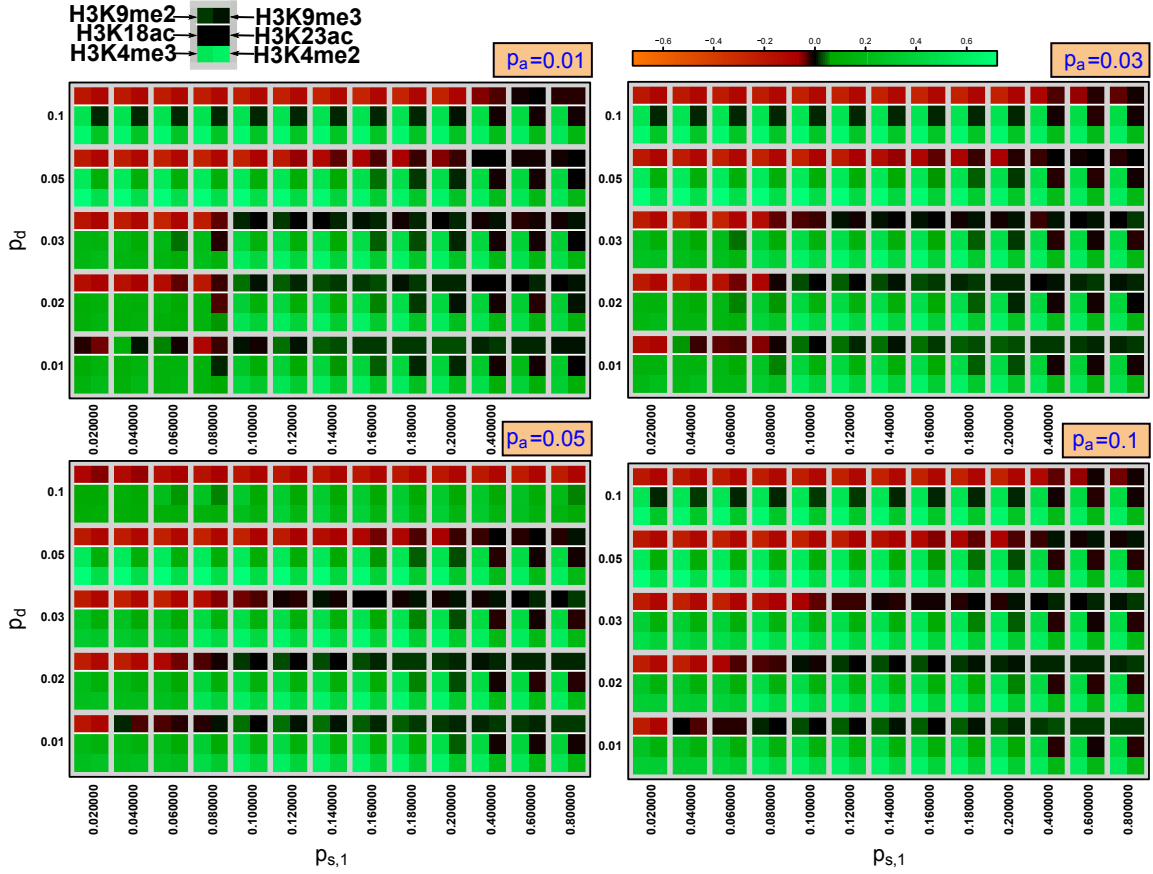}
\caption{{\bf Pearson's correlation between simulations and experiments on chromosome 1 of CD4+ cells for different values of the parameters
$p_{s,1}$,
$p_d$ and $p_a$.} A higher value corresponds to a higher correlation between the simulated chromatin distribution averaged over the last 100,000
time steps and the scores for the histone marks from the ChIP-seq analysis. The heterochromatin marks H3K9me2 and H3K9me3 were compared to the
simulated heterochromatin distribution and the marks H3K4me2, H3K4me3, H3K18ac and H3K23ac to the simulated euchromatin distribution. Hence, a good
match between simulations and experiment is obtained for all 6 fields being green. The heterochromatin marks are rather sparsely distributed and
therefore only low correlation values could be reached, especially for the H3K9me3 mark. }
\label{fig:scoreB_chr1-17}
\end{figure}

\begin{table}[!h]
\centering
\caption{\bf Parameters of the computational model giving positive correlation with the experimental data on all
simulated chromosomes. The propagation rate $p_{s,2}$ was left unchanged.}
\begin{tabular}{c|c|c|}
\multicolumn{3}{r}{}\\
parameter& function & optimal range \\
\hline
$p_a$ & association rate at nucleation sites & 0.01-0.1\\
\hline
$p_d$ & deletion rate & 0.01-0.02\\
\hline
$p_{s,1}$ & propagation rate of heterochromatin marks & $p_{s,1}=0.1-0.16$\\
\hline
$p_{s,2}$ & propagation rate of euchromatin marks & $p_{s,2}=0.1$\\
\hline
\end{tabular}
\label{table:parameters}
\end{table}

In order to find an optimal parameter set, we test the result of simulations for a large number of parameter value combinations. Association
rates at nucleation sites and deletion rates were set equal for both modifications. Furthermore, we do not allow different
parameter values for individual nucleosomes as we aim to provide a more thorough understanding of the distribution of histone modifications over
entire chromosomes. For each simulation, the measurements were carried out after the system reached its final ``stationary'' state. Experimentally
determined high nucleosome turn-over rates suggest a fast dynamics for the setting of histone marks~\cite{Deal2010,Beisel2011}. Hence, temporal
changes of the parameters are assumed to lead to fast rearrangement of the chromatin domains. Therefore, it is sufficient to compare the final states
of simulations carried out with different parameter values. 

\begin{figure}[!h]
\includegraphics[width=0.95\textwidth]{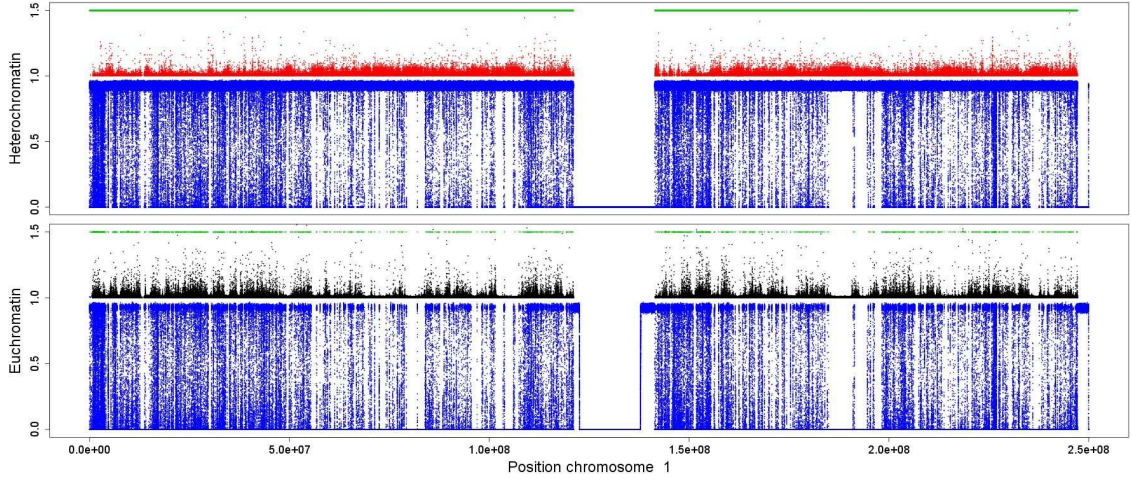}
\caption{{\bf Simulation results (blue)
and experimental data (red/black) of CD4+ T cells exhibit similar distributions for euchromatin and heterochromatin on chromosome 1.}
The red dots show both H3K9me2 and H3K9me3 marks together, i.e. plotting their scores. The black dots exhibit all scores for the marks H3K4me2,
H3K4me3, H3K18ac and H3K23ac. For each histone we depict its occupation frequency averaged over the last 100,000 time steps
after the simulations reached the relaxed state. 
The processes nucleation, propagation, deletion and competition are sufficient to reproduce the global structure of the domains on human
chromosomes.
Nucleation sites (green dots) merely function as initiators of the process whereas propagation acts as the main competitor in the system. The
blue dots show the histone mark distribution. The black and red dots correspond to the same experimental data from Figure~\ref{fig:Bmeasured}, this 
time
normalized for better visualization. Model parameters were $p_{s,1}=p_{s,2}=0.1$, $p_d=0.01$, $p_a=0.03$. }
\label{fig:CompRealChr1}
\end{figure}

In order to assess the quality of the simulation runs, we calculated the correlation values between simulations and experimental data.
The comparison yielded positive results for both types of marks, different chromosomes as well as both cell types (CD4+ T cells and HeLa cells) over a range of parameter values (Figure~\ref{fig:scoreB_chr1-17}, Figures
\ref{fig:scoreH_chr1-17} and  \ref{fig:scoreB_chrs} in File S1 and Table~\ref{table:parameters}). 
While there is a high correlation between euchromatin marks and the simulations for almost all parameter values, H3K9me2 marks agree only for
$p_{s,1}\geq 0.1$ and low values of the deletion rate, $p_d$. Euchromatin marks are known to be highly associated with CpG sites, leading to the
observed high correlation values as the nucleation sites mostly remain occupied by their respective histone marks. We could not find reasonable values
for H3K9me3 marks, probably due the specificity problems of the antibody~\cite{Duan2008}. 
For similar propagation rates $p_{s,1} \approx p_{s,2}$, the simulation results become sensitive leading to
large rearrangements of the mark distributions due to small changes of the parameters. We will further discuss the sensitivity of the system in the
next section. The correlations between simulation and experiment drop down as soon as $p_{s,1}$ becomes smaller than $p_{s,2}$. Hence,
the cells maintain the system within a state where heterochromatin marks prevail. The simulation results were similar for different association rates
at the nucleation sites, $p_a$. As consequence, while nucleation is necessary to build the core structure of the chromatin landscape, its maintenance
thereafter is mainly controlled by the strong propagation mechanism. A high degree of correlations between the simulations and the experiment was obtained for the same parameter values independently of chromosome number and even cell type. The similar results for both cell lines confirm that the majority of heterochromatin domains remain fixed even after cell differentiation, yielding a stable structure for the expression of e.g. house-keeping genes.

However, the model cannot be fitted to arbitrary histone mark distributions. Perturbations of the nucleation site distribution reduced the correlations between simulated and experimental domain distributions. We ran the simulations using nucleation sites extracted from the hg19 genome, where the genomic positions are shifted as compared to the correct hg18 (originally used in the ChIP-seq experiments). Figure~\ref{fig:hg19_nucl} shows that we were not able to find parameter values that provided well-correlating simulation results as for correctly positioned nucleation sites.

\begin{figure}[!h]
\includegraphics[width=0.95\textwidth]{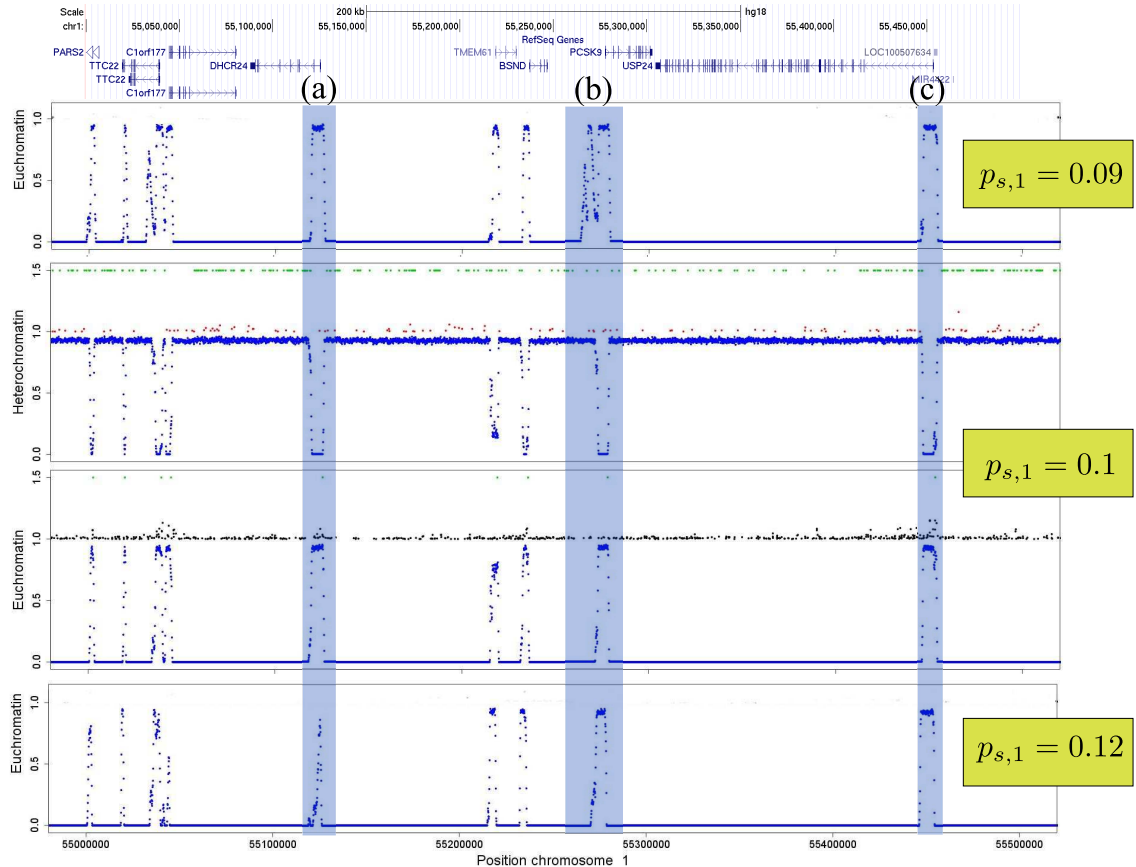}
\caption{{\bf Example for rearrangements of heterochromatin and euchromatin due to changes of the propagation rate on
chromosome 1.} We zoomed on a region of 0.5 Mbp. Red dots correspond to ChIP-seq scores of euchromatin marks and black dots to the ones of
heterochromatin marks. Green dots are respective nucleation sites and blue denotes simulations. On this scale, ChIP-seq data becomes sparse even when
taken from multiple marks. However, the heterochromatin marks mostly disappear at the euchromatin domains where the euchromatin marks reach higher
scores.  
Although this zoom was taken on an arbitrary region of the first chromosome, we see 3 different reactions of the system to alterations of the
propagation rate, $p_{s,1}$: (a) the euchromatin domain disappears for higher propagation rates; 
(b) a large euchromatin domain develops at low propagation rates; (c) the euchromatin domain remains unchanged.
(a) and (b) exhibit characteristic regions for potential gene regulations.  Refseq gene locations where adopted from
http://genome.ucsc.edu (\cite{Kent2002}). Other parameter values were $p_{s,2}=0.1$, $p_d=0.01$, $p_a=0.03$.}
\label{fig:CompRealChr1Zoom}
\end{figure}

Figure~\ref{fig:CompRealChr1} depicts one of the simulations with high correlation values to the experimental results. The formation of large
heterochromatin domains can be observed. The results are similar when simulating other chromosomes (Figures~\ref{fig:CompRealBChr2-17} and
\ref{fig:CompRealHChr2-17} in File S1). On a smaller scale, still showing the averaged state of 5000 histones, visual comparison of simulations and experimental
data becomes more difficult. Figure~\ref{fig:CompRealChr1Zoom} depicts a zoom on an arbitrary region of chromosome 1. The available ChIP-seq scores of
the histone marks are rather sparse even when considering the different marks simultaneously. 

We can produce several scenarios in our simulations that can be related to gene
activation/repression. As the model does not consider fixed borders, the dynamic nature of the model leads to rearrangements of the histone
modifications at the domain borders due to parameter changes. The borders can be constrained to narrow regions or change
along larger areas leading to broader, negotiable borders between heterochromatin and euchromatin. The range is not only controlled
by the parameter values but also in great extend by the position of the nucleation sites. 
Hence, chromatin organization can be controlled by alterations of the underlying mechanisms for histone mark propagation, nucleation, deletion and the
introduction of new nucleation sites by e.g. transcription factors. 

Figure~\ref{fig:CompRealChr1Zoom} illustrates how the landscape changes upon changes
of one of the propagation rates. We will show that the other parameters do not have an as strong impact on domain rearrangements in the next section
as well as investigate the influence of newly introduced nucleation sites. 
Region (a) of Figure~\ref{fig:CompRealChr1Zoom} provides an example for a euchromatin domain that remains stable for lower propagation rates but
almost vanishes for an increasing propagation mechanism. The opposite reaction to alterations of the propagation rate can be observed in region (b),
where the euchromatin domain nearly doubles its size. Finally, region (c) shows a case where the domain borders remain stable. A thorough look on the
distribution of the nucleation sites explains these completely different reactions of the system. The accumulation of respective nucleation sites at
opposing sides of a ``fixed'' boundary element (insulator)  allows a stable border where the fluctuations are held at a minimum in
Figure~\ref{fig:CompRealChr1Zoom}). This means that boundary elements are not blocking the propagation of e.g. heterochromatin but rather
consist of opposing nucleation sites that maintain a narrow border. While the SINEAlu elements that nucleate
heterochromatin are densely accumulated around the euchromatin domain (a), they are much less abundant in region (b) and therefore the euchromatin
domain is able to expand as soon as its propagation rate becomes larger than the one of the heterochromatic marks. 
Considerable boundary rearrangements become possible due to larger changes of the competition strength. As a consequence, heterochromatin domains
expand or shrink over larger regions leading to aberrant states (see e.g. Figure~\ref{fig:CompRealChr1Aberrant} in File S1) in the cell related to severe
diseases~\cite{Norwood2006,Cloos2006}. 

Hence, we show that specific, chromosome-wide transcription programs can be switched on and off due to small changes of the propagation rates allowing
for a effective multiple reaction to specific external stimuli. More fine-tuned gene regulation, leading to a transcriptional response of single gens
or gene groups, located within the same genomic region, can be achieved by introducing or deletion of nucleation sites and will be discussed in the
following section.

With this model based on simple assumptions for heterochromatin formation, it is possible to reproduce experimental data from humans. After
establishing this working concept, it is crucial to study general model behavior in order to understand the system's reaction to
quantitative changes of the underlying processes nucleation, propagation and deletion.

\subsection*{General model behavior, simple competition}
In order to assess the overall behavior of the model, we carried out a thorough parameter study. We focus on a system of
two competing histone marks that cannot simultaneously occupy the same histone. For most parameter combinations, stable 
chromatin domains mediated by histone modifications develop and spread from their
nucleation sites until they get stopped by competing marks. 

\begin{figure}[!h]
\psfrag{avlabel}[1][1][4]{\bf $<n_{1}>,<n_{2}>$}
\includegraphics[angle=0,width=0.6\textwidth]{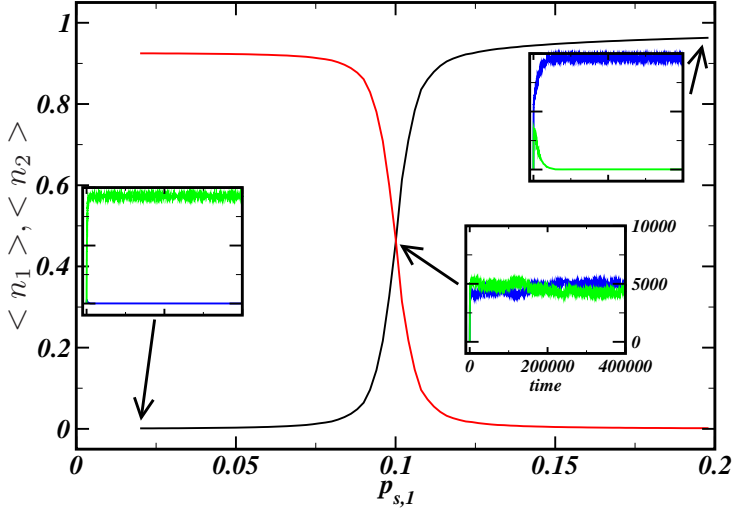}
\caption{{\bf Switch-like behavior for competing histone marks.} We take the temporal average of the number of with mark $m$ modified
histones after the simulation reached a stationary-like state, $<n_m>$, presenting now the average frequency of a histone mark. A clear transition
between two saturated states
is observed, where the number of modifications fluctuates maximally for $p_{s,1}\approx 0.1$. Inner panels: evolution of the number of modifications
for
different parameter sets. The other parameters were $N_N=100$, $p_{s,2}=0.1$, $p_{d}=p_{a}=0.01$.}
\label{fig:temp_beh}
\end{figure}

The simulation results show that small changes of the activity of molecules participating in histone
mark propagation can lead to extensive changes in the histone mark configuration. This mechanism provides an
effective on and off-switching of genomic regions. Temporal activation of one of the molecules responsible for the
propagation of a mark can result in a complete rearrangement of a chromatin region. 
Figure~\ref{fig:temp_beh} shows average frequencies for simulation runs with a different propagation rate of mark 1,
$p_{s,1}$. A sharp transition from almost full occupation by mark 2 to full occupation by mark 1 is observed. For
propagation rates $p_{s,1}\approx p_{s,2}$, both marks coexist in a stable manner. 

\begin{figure}[!h]
\psfrag{avlabel}[1][1][4]{\bf $<n_{i,m}>$}
\includegraphics[angle=0,width=0.7\textwidth]{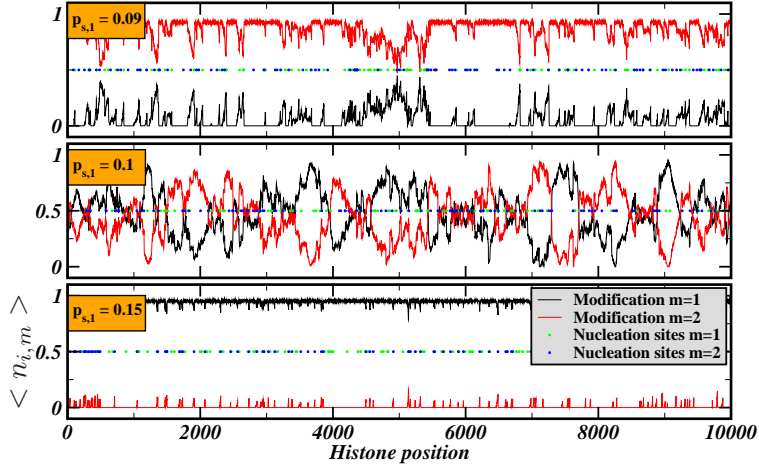}
\caption{{\bf Complete transition of the chromatin landscape for different propagation rates.} The figure shows the spatial distribution of
modifications averaged over the last iterations, $<n_{i,m}>$,  for different values of $p_{s,1}$. Despite the purely
random distribution of nucleation sites, chromatin domains form around accumulations of nucleation sites in the upper two panels. The other
parameters were $N_N=100$, $p_{s,2}=0.1$,
$p_{d}=p_{a}=0.01$.}
\label{fig:mod_distr}
\end{figure}

The spatial distribution of histone marks is depicted in Figure~\ref{fig:mod_distr}.
Different chromatin domains develop around nucleation sites. The sharp transition with
respect to different values of the propagation rate of mark 1 allows a state of distinguishable chromatin domains
near the transition point $p_{s,1}=p_{s,2}$ with negotiable borders as in the simulations presented in the previous section.
The transition becomes smoother for a larger number of nucleation sites, $N_N$, as well as for larger association rates,
$p_a$, or smaller $p_d$ (see Figures~\ref{fig:Nm_diffNN_pa} and ~\ref{fig:Nm_diffpd} in File S1). Generally, a behavior similar to
phase transitions in physical systems can be observed with $p_{s,1}$ or $p_{s,2}$ as control parameter. The strong
increase of the frequency fluctuations near the transition point ($p_{s,1}=p_{s,2}$) supports this finding. This behavior allows high
sensitivity and fast response of the system to stimulation.
A non-random distribution of nucleation sites can maintain domains stable against these fluctuations, allowing different responses of the domains to changes of the propagation rate as could be seen in Figure~\ref{fig:CompRealChr1Zoom}. Therefore, the system allows a variety of different regulatory responses induced by a simple control mechanism.

Sensitive global switch-like behavior occurs only for simulation scenarios where propagation rates become altered. For different
nucleation rates (Figure~\ref{fig:Nm_ch_pa} in File S1) and different deletion rates (not shown), the system exhibits no drastic
changes in its behavior and there is no peak in the fluctuations at the transition point. This result is important for the understanding of switching
of large chromatin domains. While single genes and smaller genomic regions are most likely switched on/off through
alterations in the nucleation sites by e.g. transcriptions factors, a fast and
complete response of the system can only be achieved by changing the activity of the agents involved in the propagation of histone modifications.
The processes behind nucleation and deletion of these histone marks should play an indirect role. Their function relies in the
``regulation of the regulation'' by changing the sensitivity.  
Hence, as small changes of the propagation rates imply large-scale expansion/shrinking of chromatin regions, the cell should incorporate
regulatory elements within the machinery behind the propagation of e.g. heterochromatin (interaction between HP1, Suv39h1 and deacetylases)
and H3K27me3 (Polycomb multi-protein complexes). Indeed, the activity of these protein complexes might be fine-tuned by post-translational
modifications of certain Polycomb proteins and HP1~\cite{Hatano2010,Kang2010,LeRoy2009,Beisel2011}. While HP1-controlled heterochromatin domains marked by H3K9me2 and H3K9me3 generally maintain their structure for different cell types, Polycomb complexes and H3K27 methylations play an important role in cell differentiation. We propose that fine-tuning of the activity of the Polycomb machinery might be crucial to switch respective chromatin domains in order to guarantee transcription of cell type specific genes. Nucleation sites for H3K27 methylation domains are still not well-understood and we intend to incorporate the formation of H3K27 methylation domains as soon as detailed information about the nucleation sites becomes available.

\begin{figure}[!h]
\includegraphics[angle=270,width=0.7\textwidth]{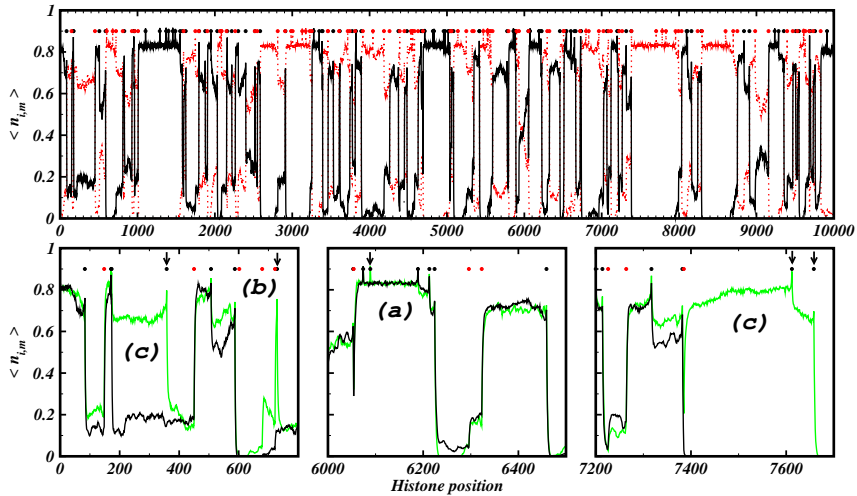}
\caption{{\bf New nucleation sites can lead to different effects.} Upper panel: Black and red lines denote the spatial distribution of histone mark 1
and 2, respectively. The 90 (100) nucleation sites for mark 1 (2) are shown as black (red) points. Lower panels: Zooms on the spatial distribution of
mark 1 for 90 nucleation sites (black line) and 100 nucleation sites (green line). The 100 nucleation sites of mark 2 remain unchanged. The arrows tag
new inserted nucleation sites of mark 1. Three different nucleation effects can be observed: (a) no change ; (b) narrow spike around nucleation site ;
(c) activation of large region. The simulation parameters were $p_{s,2}=p_{s,1}=0.1$, $p_{d}=0.02$ and $p_{a}=0.1$.}
\label{fig:NuclSEffect}
\end{figure}

Locally directed regulation of transcription can be achieved by small changes within the nucleation sites. The position of a new nucleation
site as well as its surrounding ones are main factors for the specific response of the system. Figure~\ref{fig:NuclSEffect} illustrates
that these responses can have different regulatory effects on the transcription of genes. We compare the chromatin mark distributions of two
simulations where we vary the number of nucleation sites of mark 1 from 90 to 100 while the 100 nucleation sites of mark 2 remain
unchanged. We can distinguish three different situations with respect to the introduction of a new nucleation site: (a) the chromatin domain remains
stable and unchanged; (b) The new nucleation site leads to a small peak around its position leading for instance to the activation/silencing of a
single gene; (c) the chromatin domain expands drastically over are large region and therefore has potentially impact on the transcription of multiple
genes. 
Hence, long-range effects are possible by small changes of the distribution of nucleation sites. Therefore, (de-)activation of nucleation sites can
have drastic consequences to a cell. \cite{Finelli2012} showed experimentally that chromosomal translocation can lead to a long-range position-effect
variegation. The erroneous recombination of chromosomes leads to an untypical configuration of the nucleation sites around the recombination site
that might have strong impact on the transcriptional program of nearby genes. 

\section*{Discussion}
\emph{Minimal model for chromatin domain formation.}
The simulations show that it is possible to reproduce the main aspects of experimentally observed heterochromatin and euchromatin marks on human
chromosomes with a model based only on the principal mechanisms of chromatin domain formation. Although the \emph{in vivo} system involves a
myriad of molecules that play a role in the regulation of these patterns, the simple approach based on nucleation,
propagation and competition was sufficient to obtain a stable system of coexisting marks. 

\noindent
\emph{Active competition between chromatin domains instead of fixed boundary elements.}
This scenario works only when
competing marks are actively maintained through propagation processes. Passive insulators cannot stop heterochromatin from
completely shutting down the chromosomes. Additionally, active competition results in negotiable borders between
different modification domains, allowing for sensitive regulation of broad domains that might contain multiple genes. Moreover, this scenario
does not require specific insulators/boundary elements preventing e.g. the propagation of chromatin by blocking the propagating machinery as already
suggested by~\cite{Hathaway2012}. The stability of these elements can be explained by proper accumulation of the nucleation sites at opposing sides of
two competing marks. Fixed boundaries are maintained by strong nucleation effects near the border, similar to the model of presented in
ref.~\cite{Wang2012} where a boundary element locally recruits histone-modifying enzymes and transcription factors. 

\noindent
\emph{Distribution of nucleation sites defines domain formation and regulatory features.} Transcriptional regulation
is mainly controlled by the positional arrangement of the nucleation sites on the genome. 
 Thus the specific distribution of nucleation sites plays
a large role for the transcriptional program of a cell. 
Particular arrangements of nucleation sites promote or inhibit gene activation/repression due to changes of the propagation machinery or introduction of nucleation sites by e.g. transcription factors. Mutations of until now mostly disregarded parts of the DNA could have strong impact on
an organism leading to dysfunctional states like diseases.

\noindent
\emph{Propagation of chromatin marks as global actor.} Furthermore, molecules controlling propagation seem to be an important factor for
regulation of these patterns,
able to rapidly switch large chromatin domain regions. Whereas the other factors leading to nucleation and deletion of marks
remain crucial for domain formation and single-gene switching, small changes in the propagation rates can lead to rearrangement of the
transcriptional program. The observed behavior allows a fast response of the system to external stimuli by changing the activity
of entire gene regions. The identification of several post-translational modifications of the proteins involved in the propagation
of H3K9 and H3K27 methylation~\cite{Hatano2010,Kang2010,LeRoy2009,Beisel2011} experimentally support this important role of
the propagation machine. 

\noindent
\emph{New nucleation sites can switch on/off large gene groups.} Transcriptional changes of single genes of gene groups can be achieved by small changes in the distribution of the
nucleation sites. Even the introduction of only one new nucleation site can lead to long-range effects and therefore function as switches for entire
gene groups. This effect can be observed experimentally, leading to long-range position-effect variegation due to recombination of the wrong
chromosomes~\cite{Finelli2012}.

\noindent
\emph{Dynamics of chromatin domain formation complicates simple experimental recognition of nucleation sites.} It is crucial to bear in mind
that both nucleation sites and propagation are
responsible for the structure of chromatin domains. Therefore, studies that simply measure correlations between genomic motifs and histone
modifications will not succeed to unravel their relation. Experiments should combine information about multiple histone
modifications and their position. The liaison of two already widely applied experimental techniques has the power to
achieve this task. By combining mass spectrometry identifications of multiple modifications and ChIP-seq experiments
revealing the position of simple modifications, we should gather sufficiently specific data. Furthermore, the
identification of interaction rules and nucleation sites requires sophisticated bioinformatics tools to extract the
relevant information~\cite{Sidoli2012}.

% \add{The main results were summarized in Table~}\ref{table:results}.
\section{Conclusions}

Our study shows that the complex regulatory machinery involved in the formation of histone patterns can be
modeled and simulated by a simple system using a series of basic rules. Two key players for the rearrangement of chromatin domains were identified,
namely the propagation rate and nucleation sites, that should play a crucial role in switching between different genetic programs in the cell. Even
small alterations of the propagation mechanism responsible for histone mark spreading can lead to global changes. We suggest, that experiments focus
on the key molecules responsible for
mark propagation as they seem to have major regulatory functions. On the other hand, insertions or deletions of single nucleation sites can result in
switching on/off large groups of colocated genes. The strength of the effect depends mainly on the distribution of nearby
nucleation sites. Hence, we suggest that including further knowledge about this distribution may add highly relevant information when studying the
impact of transcription factors on nearby genes.

The presented system may serve for the construction of a general theory for the language of histone marks related to domain formation. It can be
used in detailed studies of the regulation of single gene groups by incorporating further parameters such as transcription factors and their
interaction with the DNA. Moreover, the model can be extended to simulate cross-talk between multiple histone marks and possibly reveal the mechanisms
responsible for the formation of Polycomb domains, known to play a crucial role in cell differentiation.

\section*{Methods}
\subsection*{Data analysis}
{\bf Chip-seq:} 
We analyzed the raw data files for CD4+ T cells and HeLa cells with the CCAT tool (version 3.0)~\cite{Xu2010}. The goal was to reveal
the overall structure of the histone mark distribution without requiring high-confidence scores for every single measurement. Therefore, we changed
the default parameters to a minimal count of 2 hits in a sliding window of 500 bp and lowered the minimal significant score from 3 to 2.

The data for CD4+ T cells was adapted from~\cite{Barski2007a,Wang2008} and the raw bed-files were download from the provided internet sites. The
available control file was used as control library. The data for HeLa cells was originally generated by refs.~\cite{Vermeulen2010,Liu2010,Qi2010} and
the bed files were downloaded using the CistromeMap interface (http://cistrome.org). We used the control library from~\cite{Vermeulen2010}. All
positions were aligned to human reference genome hg18.

\noindent
{\bf Correlations between histone modifications: }
Each chromosome was divided into bins of 100 kbp. For each bin, we took the sum of all scores of the corresponding histone modification. The sums
were used to calculate Pearson's correlation between two modifications. We ignored bins with one of the two sums equal to zero in order to avoid
effects coming from a low antibody specificity. 

\subsection*{Computational model}
The rules of the computational model to simulate the distribution of histone marks were implemented as stochastic processes. These processes
are expressed as rates, giving the probability to change per time step and histone. Constant stochastic rates for the setting and deletion of histone
marks assume that the number, activity and specificity of the involved enzymes does not change within a simulation run. The reaction of the
system to changes of parameter values is evaluated by comparing different simulation runs with respective values.\\
\noindent
{\bf Histone chain:} We simplify the complex chromatin structure by neglecting the exact positions of the
nucleosomes. A chromosome is modeled by a one-dimensional chain consisting of $N_H$ nucleosomes containing histone H3,
allowing only nearest-neighbor interactions. The effect of the structural conformation of the chromatin fiber and resulting long-range interactions
could be simulated by introducing an additional source of noise to the system. We neglect this effect in order to minimize the number of parameters
and also because we think that such an extension does not lead to drastic changes in the simulations.  When simulating a dynamical epigenetic
landscape on human chromosomes,
one histone H3 molecule is located for each 100 bp of DNA, rather than two molecules per $\approx$ 200 bp. \\
\noindent
{\bf Nucleation sites: } The model considers two scenarios for nucleation site placement: (a) For simulations of
human chromosomes, the genomic positions of the nucleation sites on the chromosomes are inserted using the data obtained
from RepeatMasker~\cite{Smit1996} and the UCSC Genome Browser~\cite{Fujita2011}. 
Specifically, we downloaded
hg18 from repeatmasker.org (repeat library 20080120) and
CpG sites were extracted from the table browser at http://genome.ucsc.edu/, searching for cpgIslandExt in hg18 version 2006.
hg19 versions were downloaded from repeatmasker.org (repeat library 20120124) and the UCSC Genome
Browser (hg19 version Feb. 2009). As the here analyzed ChIP-seq data is based on hg18, we used the therefore incorrect hg19 nucleation sites to check the performance of the simulations on a perturbed system.
We assumed that heterochromatin marks and euchromatin marks are nucleated by
SineAlu sequences and by CpG sites, respectively. (b) For the study of the general model, we fix the number of
nucleation sites $N_N$ that then will be randomly set on the histone chain. This number is the same for different marks,
i.e. each modification type will be fed by the same number of nucleation sites.
During the simulations, histones get modified with a new mark of type $m$ with rate $p_{a,m}$ on each nucleation site, i.e. a histone gets
modified if a random number between 0 and 1 is smaller than $p_{a,m}$\\
\noindent
{\bf Propagation of marks: }
Unmodified histones become modified either directly in the case of a nucleation site or by propagation of the mark from
a neighboring histone. The rate for a mark $m$ to propagate to a neighboring site corresponds to $p_{s,m} $. This new
mark can only be set when the site is not already occupied by a competing mark.\\
\noindent
{\bf Deletion: }
Histone modifications are generally reversible. Demethylases delete histone marks and maintain the dynamics of the
histone mark distribution. In our model, the marks are erased with the rate $p_{d,m}$ being in our case the same for
different modification types, i.e. we set $p_{d,m}=p_d$. The parameters for histone mark association on nucleation sites, $p_{a,m}$, and the rates for
mark deletion, $p_d$, are set
to lower values than the rates for mark propagation. Small $p_{a,m}$ values allow the situation
that nucleation sites become occupied by a competing mark, being essential for switching behavior.\\
\noindent
{\bf Simple competition: }
The processes nucleation, propagation and deletion are essential to create a dynamic state that self-organizes in
domains of distinct histone modification marks. We do not allow multiply
modified histones. As a consequence, histone modification domains form strictly non-overlapping
regions.\\
\noindent
{\bf Time evolution: }
After initialization with nucleation sites, the initially completely unmodified histone chain is updated at each time step $N_H$ times
according to the following updating scheme:
(i) Fetch random histone, (ii) if the histone is modified, the mark is deleted or spread to one of the neighboring sites
with the corresponding rates, (iii) if nucleation site, put mark with corresponding rate.
As we do not have experimental values for the different rates used in the model, the length of a time step is defined relatively to
one of the propagation rates, i.e. all rates are considered relative to $p_{s,m}$\\
{\bf Score for comparison with experimental data:}
Experimental data coming from ChIP-seq experiments are noisy and can be quite sparse leading large amounts of missing signals. Therefore, we compare
simulation and experiment only at the sites where a modification was detected. In detail, the average frequency $f_{j,m}$ of
histone mark $m$ on site $j$ is compared to the value of the significance score $c_{j,m}$ obtained with the CCAT tool~\cite{Xu2010}. Therefore, we
estimate similarity between experiment and simulations by calculating Pearson's correlation between simulations and ChIP-seq measurements for the
marks H3K9me2, H3K9me3, H3K4me2, H3K4me3, H3K18ac and H3K23ac,
\begin{equation}
 C =  \frac{<(c_{j,m}-<c_{j,m}>) (f_{j,m}-<f_{j,m}>)>}{\text{s.d.}(c_{j,m}) ~\text{s.d.}(f_{j,m})}~,
\end{equation}
where $<>$ denotes the average over all nucleosomes that have been associated with experimental values and $\text{s.d.}$ is the standard deviation.
$C=0$ corresponds to no similarity between simulation and experiment.
\bibliography{histone}
\bibliographystyle{plos2009}

% \section*{Acknowledgments}
%  VS and ONJ acknowledge generous financial support from the Danish Council for Independent Research, Natural Sciences (FNU), and the Danish National
% Research Foundation (Center for Epigenetics, grant number DNRF82).

\clearpage

% \begin{table}[!h]
% \centering
% \caption{\bf Main results of the minimal model.}
% \begin{tabular}{|c|c|}
% \multicolumn{2}{r}{}\\
% result & evidence \\
% \hline
% Minimal model can reproduce chromatin distribution & High correlations with CHiP-seq data\\
% No specific boundary elements required & Instead competition and active recruitment of opposing marks~\cite{Hathaway2012,Wang2012}\\
% Regulatory response defined by nucleation site distribution & Long-range position effect variegation~\cite{Finelli2012}\\
% Histone PTM propagation as regulatory mechanism & Many PTMs on involved proteins~\cite{Hatano2010,Kang2010,LeRoy2009,Beisel2011}\\
% New nucleation site can switch large gene groups& Long-range position effect variegation~\cite{Finelli2012}\\
% \hline
% \end{tabular}
% \label{table:results}
% \end{table}

\FloatBarrier
\renewcommand{\thefigure}{S\arabic{figure}}
\setcounter{figure}{0}
\begin{figure}[h!]
\centering
\includegraphics[angle=0,width=1.0\textwidth]{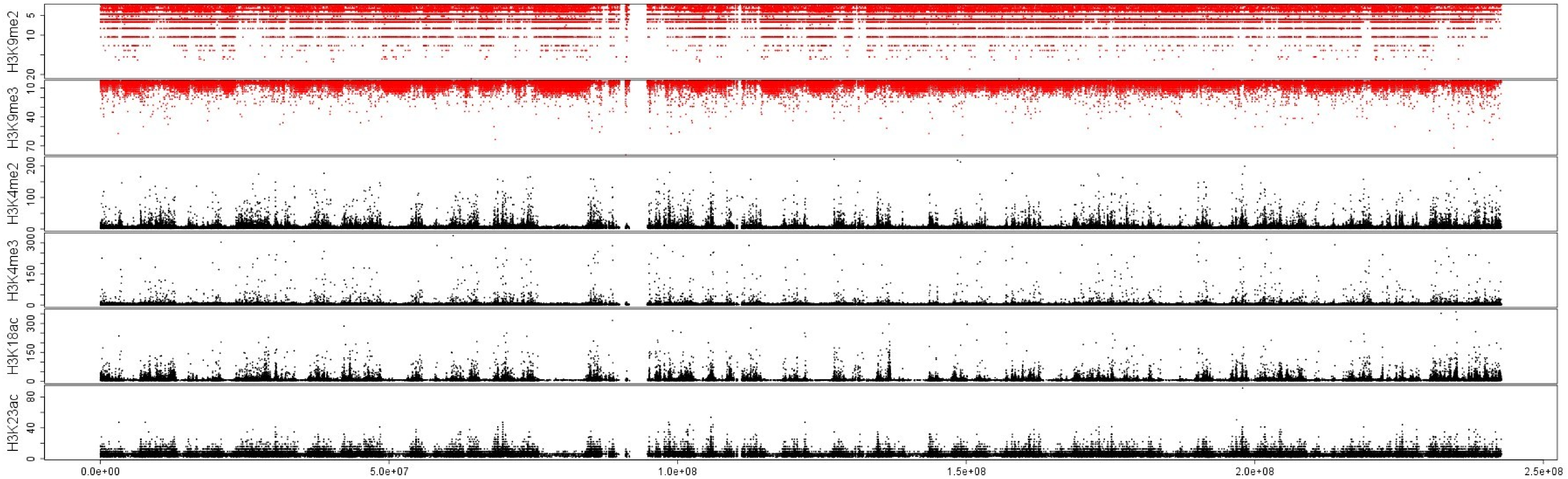}
\caption{Distribution of histone modifications measured with ChIP-seq experiments on human CD4$+$ T cells for chromosome 2. }
\label{fig:Bmeasured2}
\end{figure}

\begin{figure}[h!]
\centering
\includegraphics[angle=0,width=1.0\textwidth]{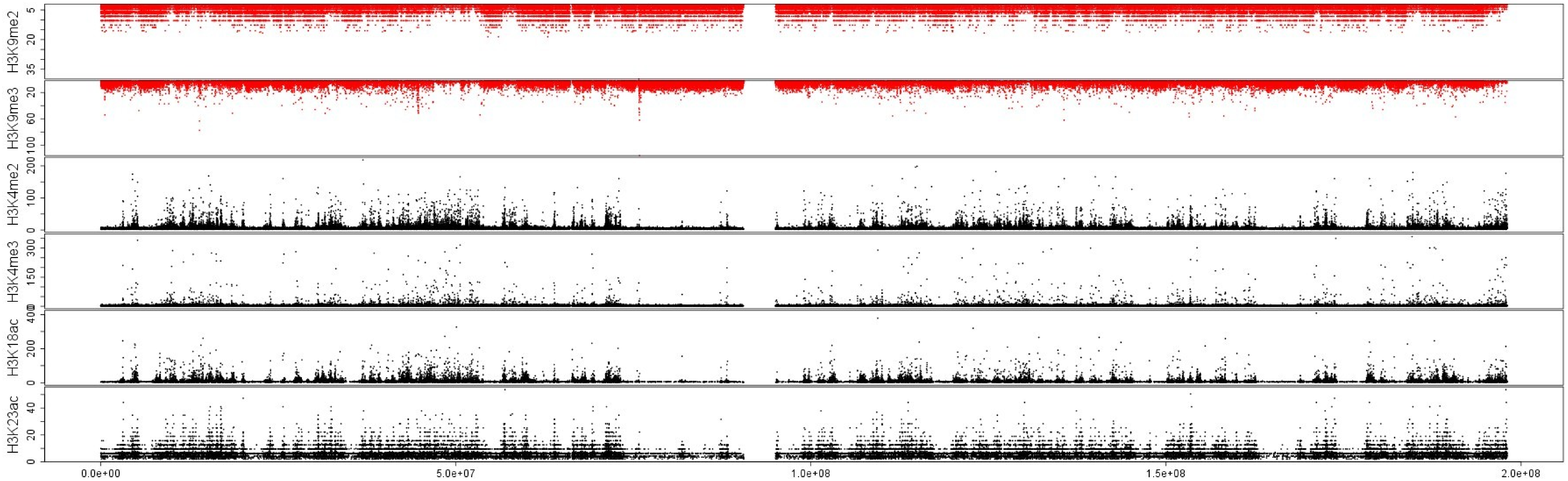}
\caption{Distribution of histone modifications measured with ChIP-seq experiments on human CD4$+$ T cells for chromosome 3. }
\label{fig:Bmeasured3}
\end{figure}

\begin{figure}[h!]
\centering
\includegraphics[angle=0,width=1.0\textwidth]{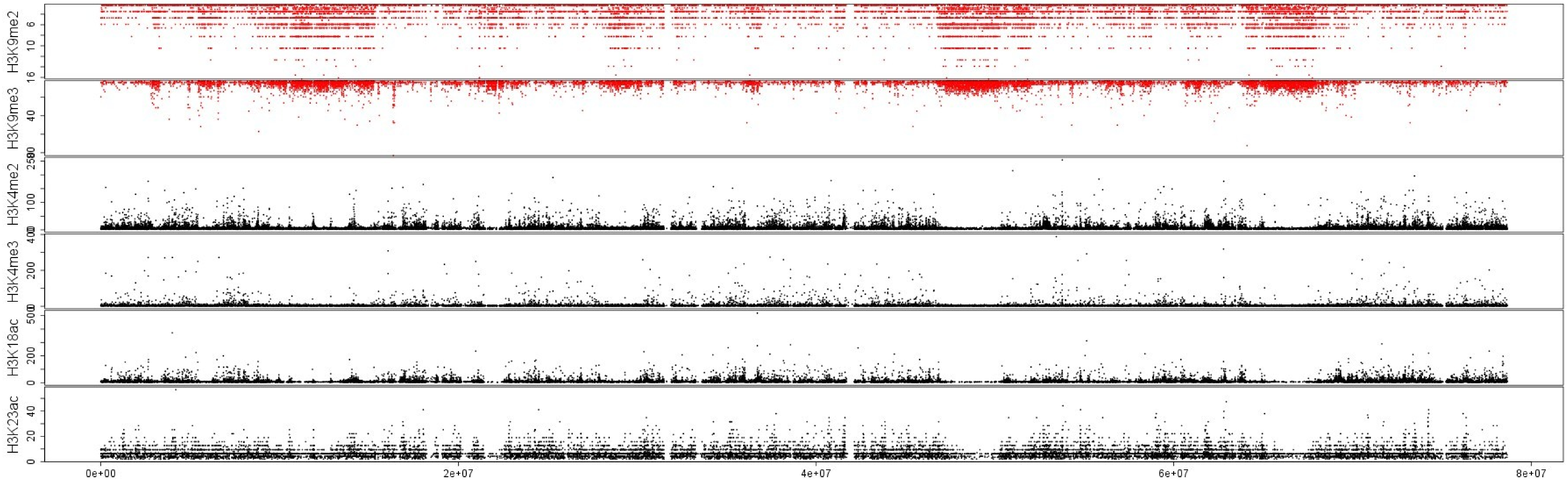}
\caption{Distribution of histone modifications measured with ChIP-seq experiments on human CD4$+$ T cells for chromosome 17. }
\label{fig:Bmeasured17}
\end{figure}

\begin{figure}[h!]
\centering
\includegraphics[angle=0,width=1.0\textwidth]{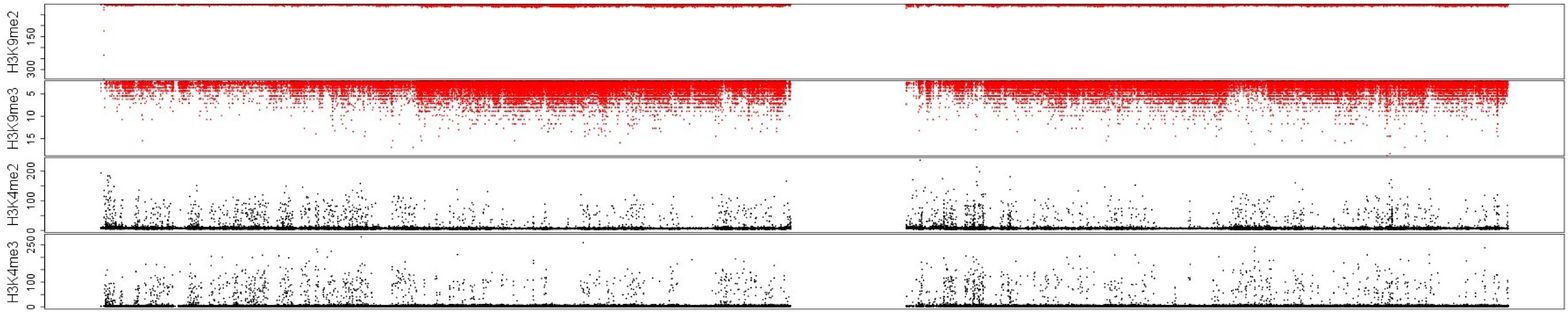}
\caption{Distribution of histone modifications measured with ChIP-seq experiments on human HeLa cells for chromosome 1. }
\label{fig:Hmeasured1}
\end{figure}

\begin{figure}
\centering
\includegraphics[angle=0,width=\textwidth]{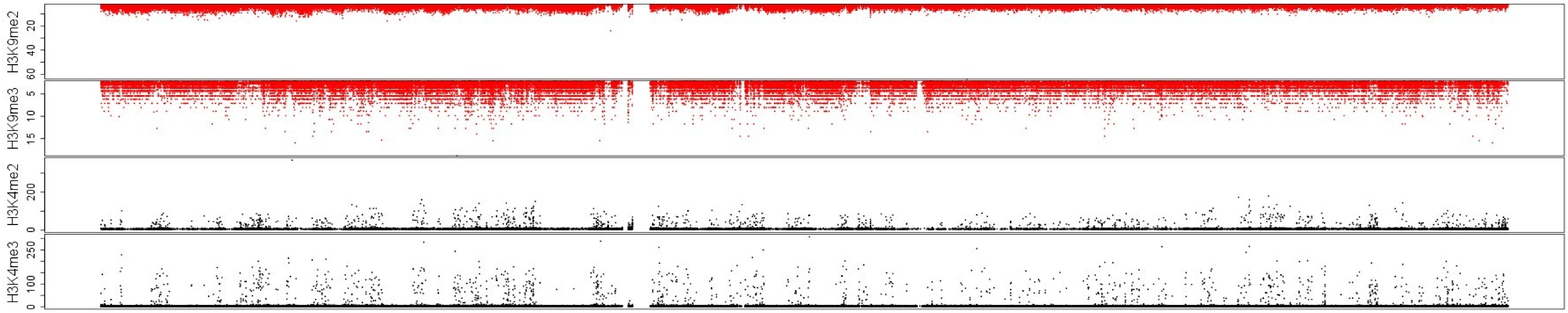}
\caption{Distribution of histone modifications measured with ChIP-seq experiments on human HeLa cells for chromosome 2. }
\label{fig:Hmeasured2}
\end{figure}

\begin{figure}[h!]
\centering
\includegraphics[angle=0,width=\textwidth]{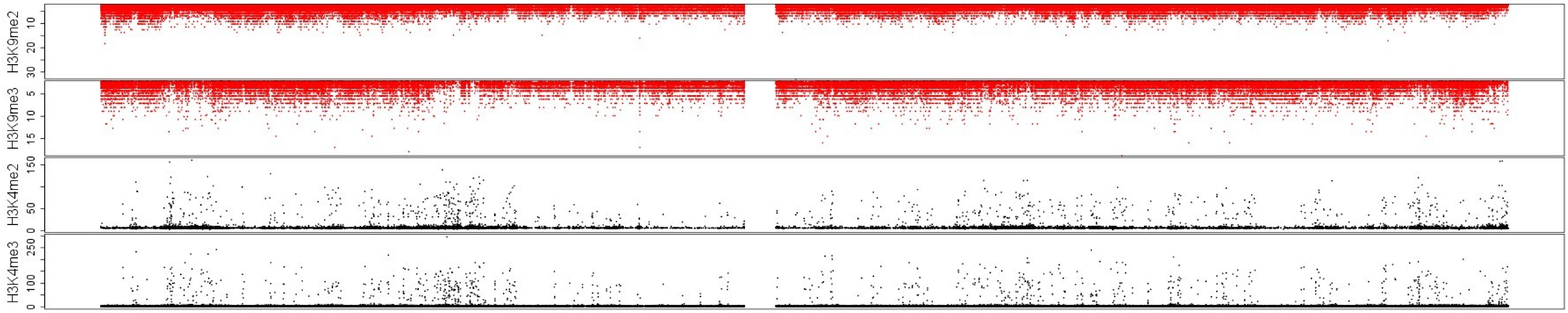}
\caption{Distribution of histone modifications measured with ChIP-seq experiments on human HeLa cells for chromosome 3. }
\label{fig:Hmeasured3}
\end{figure}

\begin{figure}[h!]
\centering
\includegraphics[angle=0,width=\textwidth]{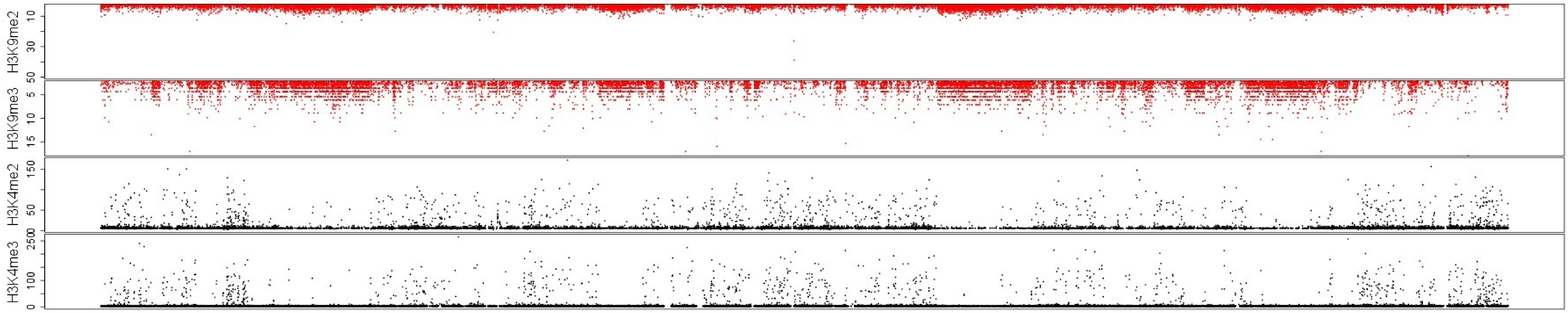}
\caption{Distribution of histone modifications measured with ChIP-seq experiments on human HeLa cells for chromosome 17. }
\label{fig:Hmeasured17}
\end{figure}

\begin{figure}[h!]
\includegraphics[angle=0,width=0.95\textwidth]{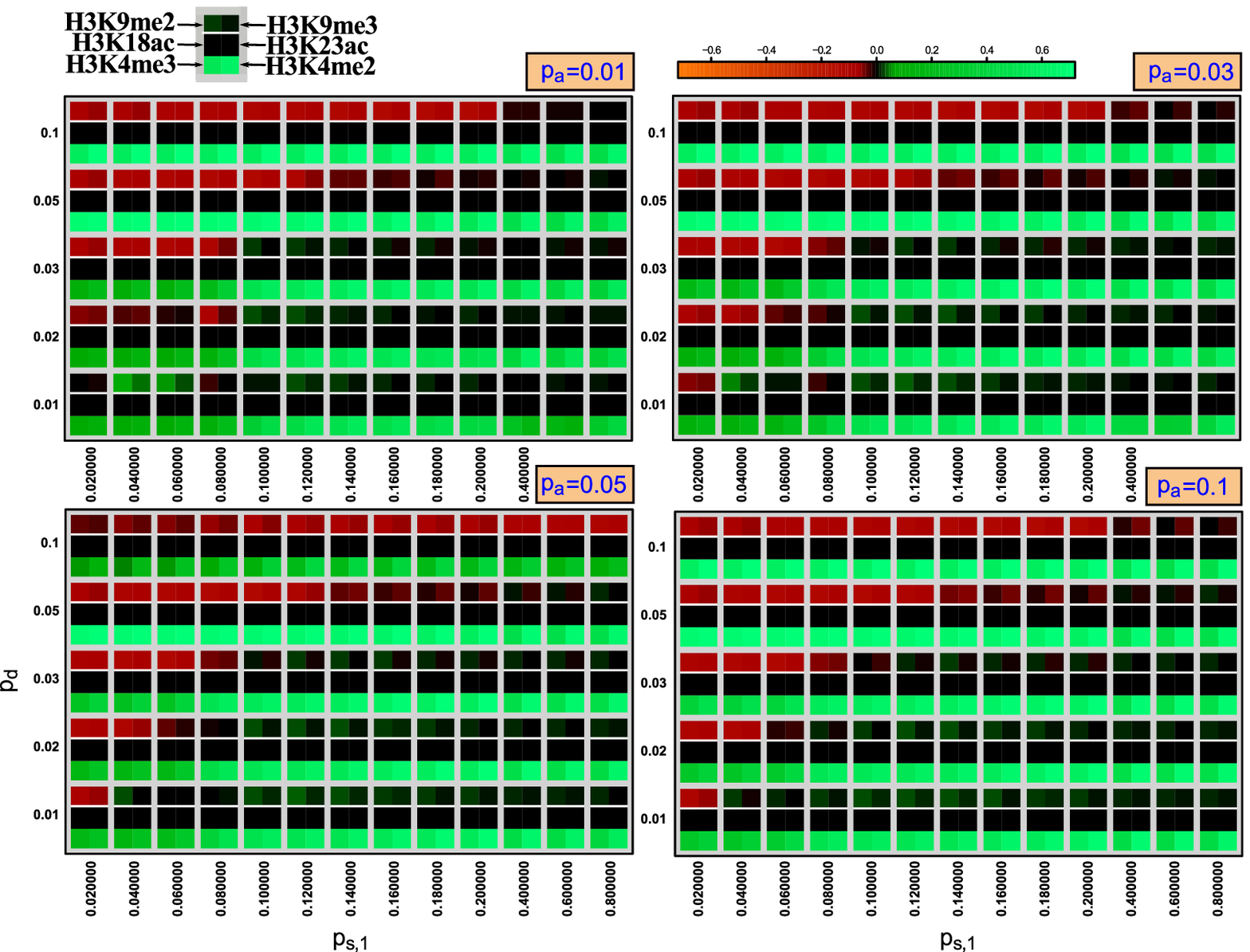}
\caption{Pearson's correlation between simulations and experiments on chromosome 1 on HeLa cells for different values of the parameters $p_{s,1}$,
$p_d$ and $p_a$. No data was available for the marks H3K18ac and H3K23ac (fields remain black). The results are almost identical to the ones for CD4+
cells.  }
\label{fig:scoreH_chr1-17}
\end{figure}

\begin{figure}[h!]
\includegraphics[angle=0,width=0.95\textwidth]{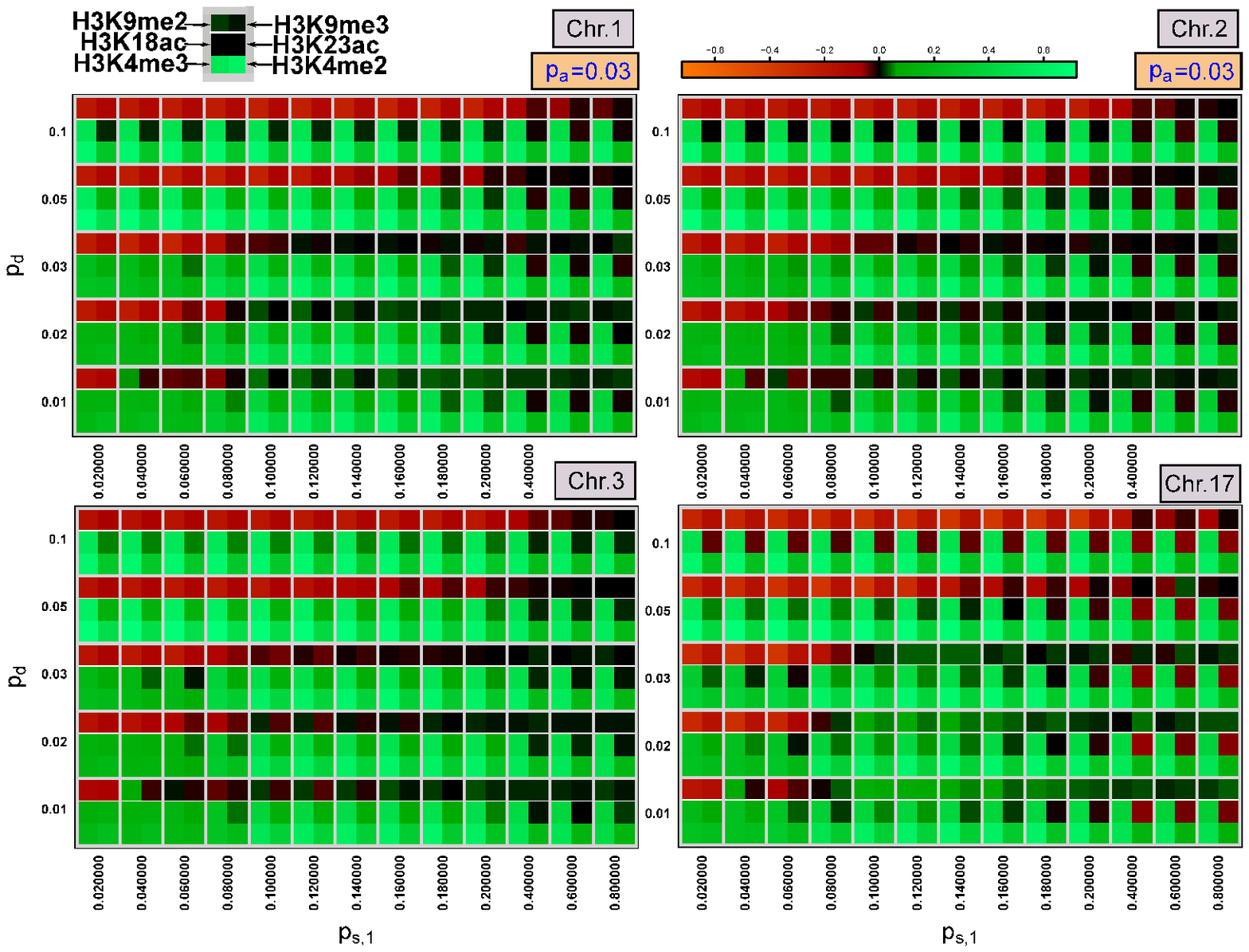}
\caption{Pearson's correlation between simulations and experiments on chromosomes 1-3 ad 17 on CD4+ cells for different values of the parameters 
$p_{s,1}$,
$p_d$ and $p_a=0.03$. The results are very similar between the different chromosomes.  }
\label{fig:scoreB_chrs}
\end{figure}

\begin{figure}[h!]
\includegraphics[angle=0,width=0.95\textwidth]{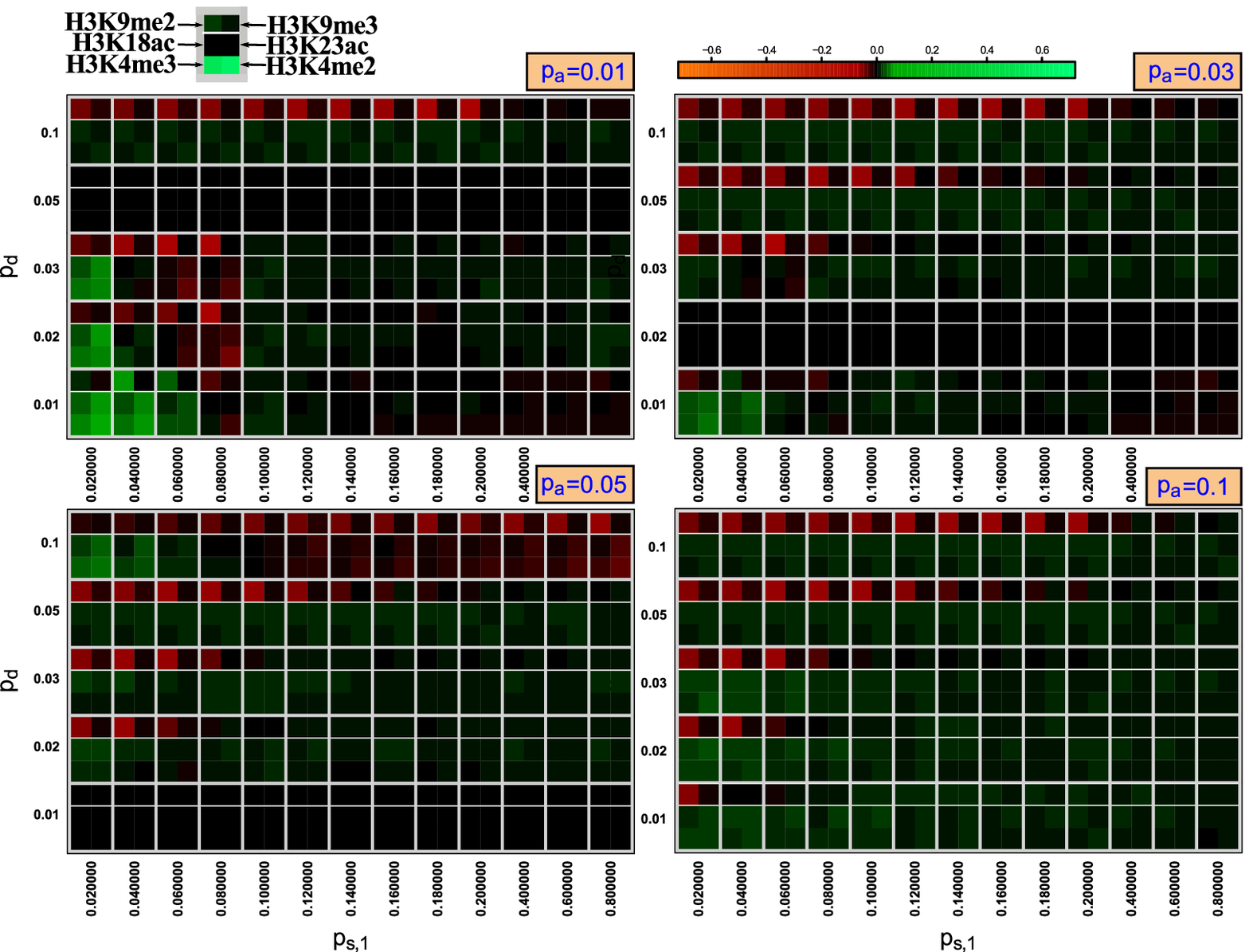}
\caption{Pearson's correlation between simulations and experiments on chromosome 1 on CD4+ T cells for different values of the parameters $p_{s,1}$, 
$p_d$ and $p_a$. We used wrong coordinates for the nucleation sites based on human genome hg19 instead of hg18. Resulting correlation values are much 
lower than for correct nucleation site positions. Hence, these slightly differently positioned sites impede accurate reproduction of the chromatin 
domains for all parameter values.}
\label{fig:hg19_nucl}
\end{figure}

\begin{figure}
\centering
\includegraphics[width=0.9\textwidth]{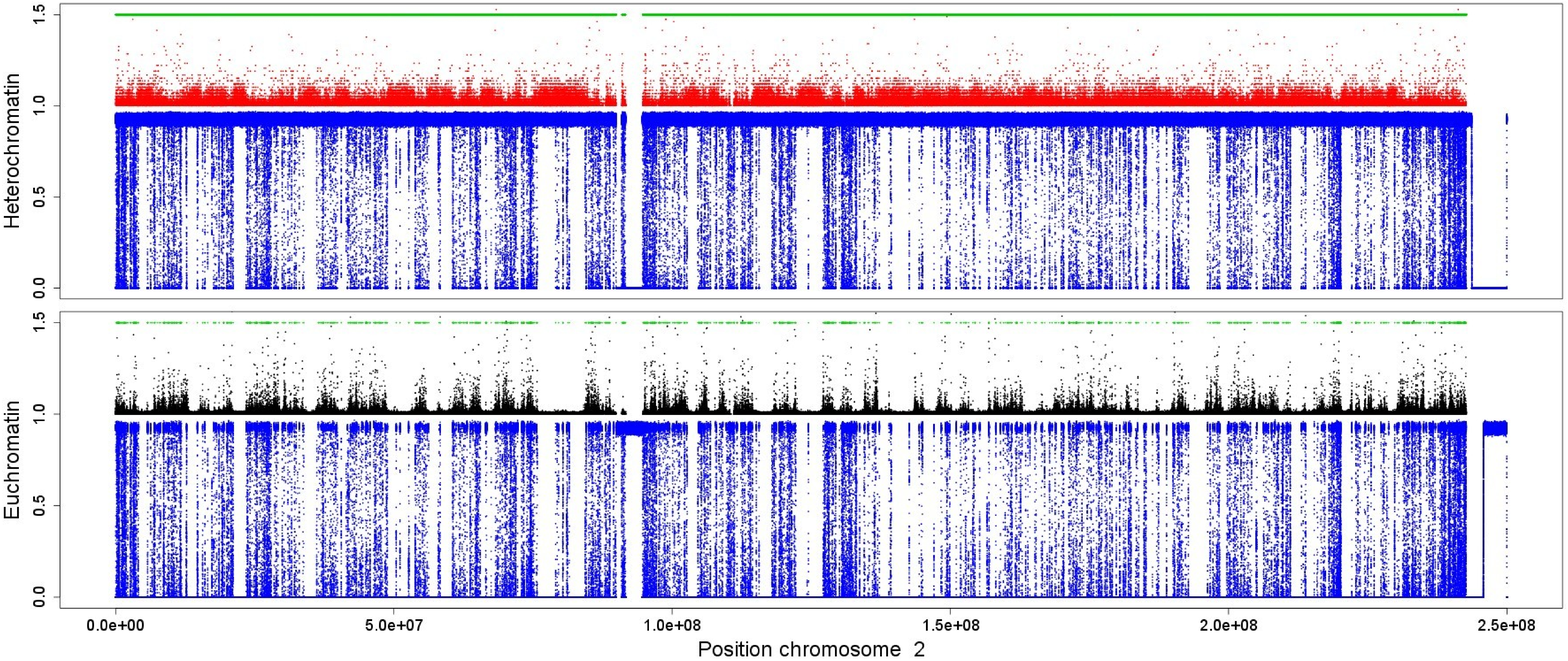}
\includegraphics[width=0.9\textwidth]{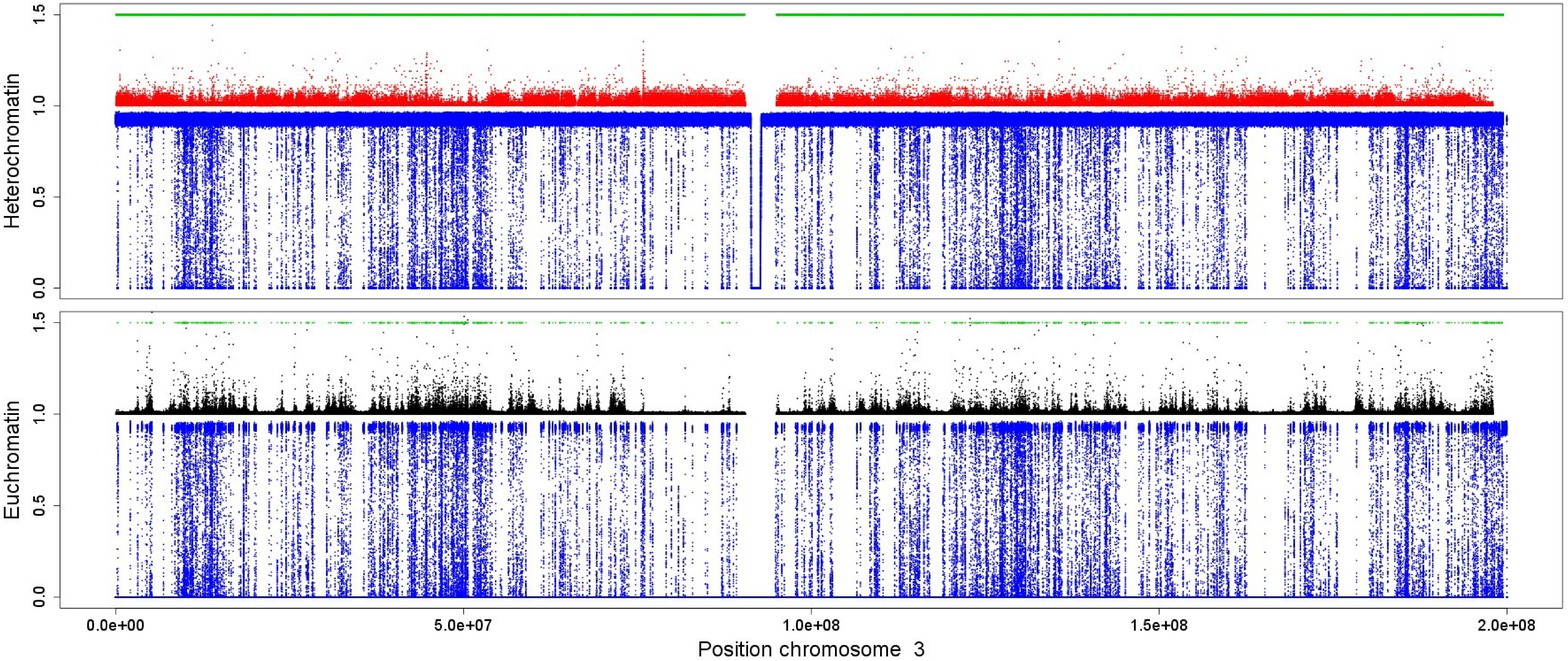}
\includegraphics[width=0.9\textwidth]{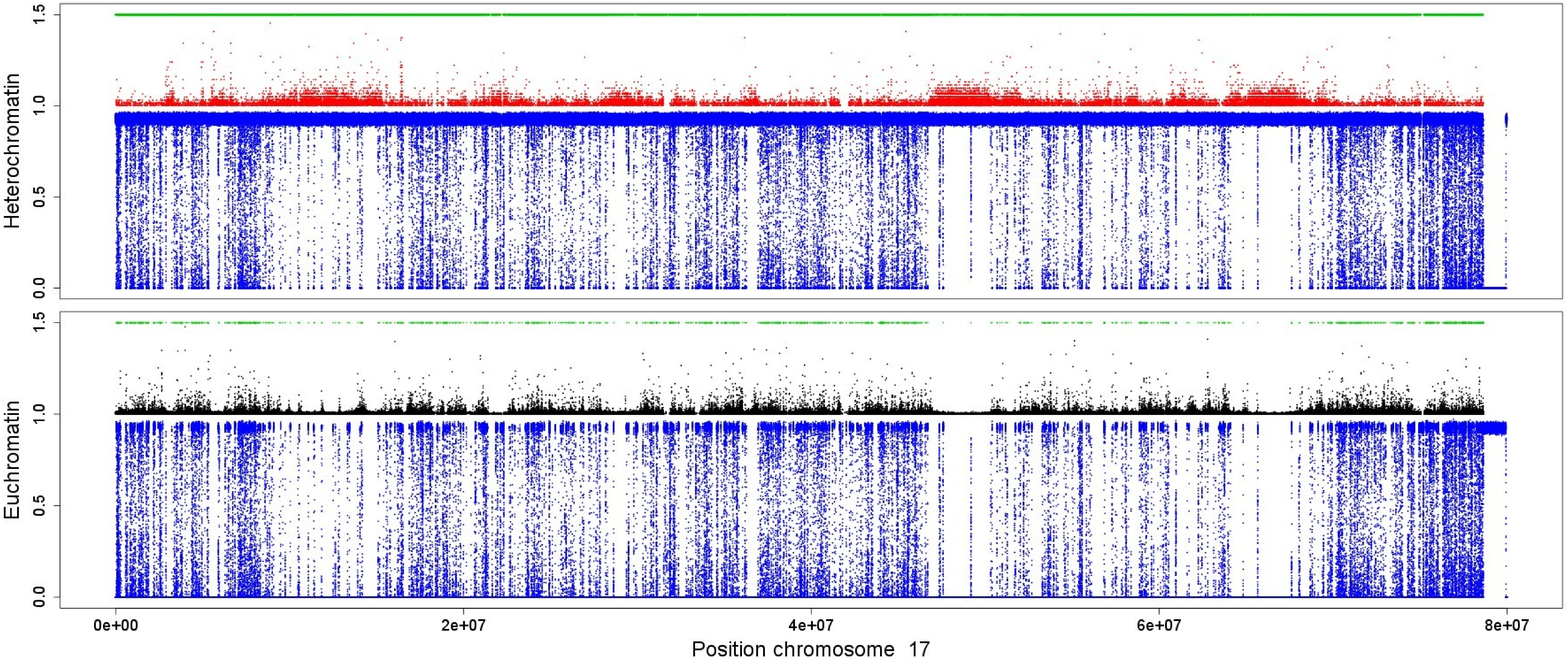}
\caption{Comparing the simulation results for chromosomes 2,3 and 17 to the CD4+ T cells data set. Parameters are the same as in
Fig.~5.}
\label{fig:CompRealBChr2-17}
\end{figure}

\begin{figure}
\centering
\includegraphics[width=0.8\textwidth]{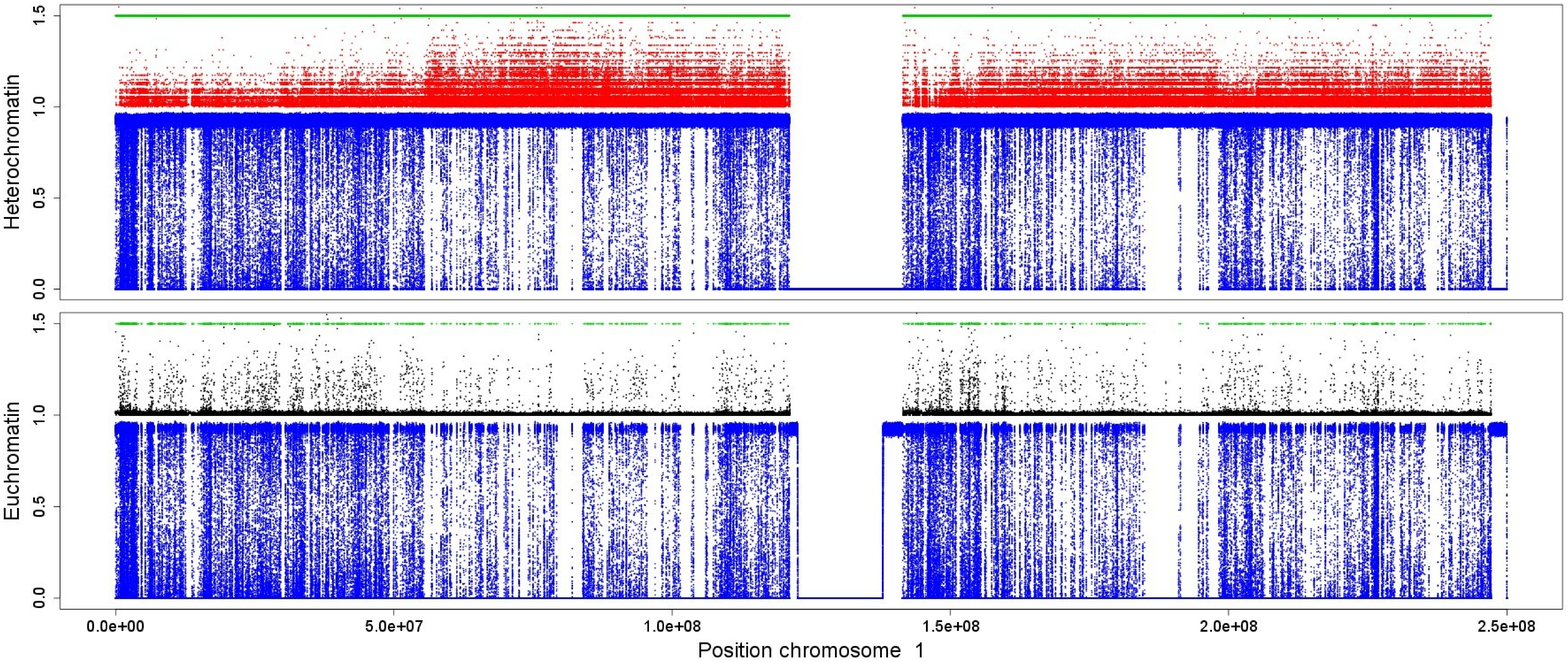}
\includegraphics[width=0.8\textwidth]{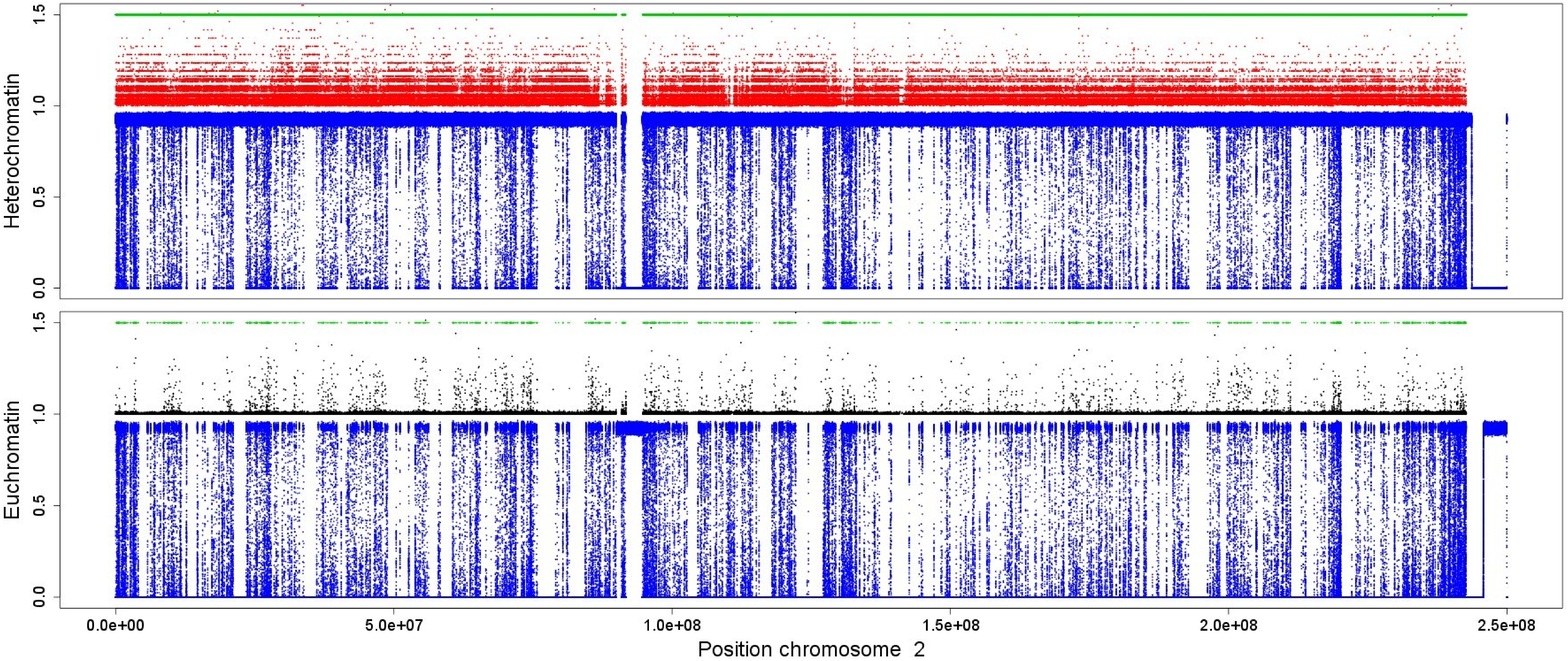}
\includegraphics[width=0.8\textwidth]{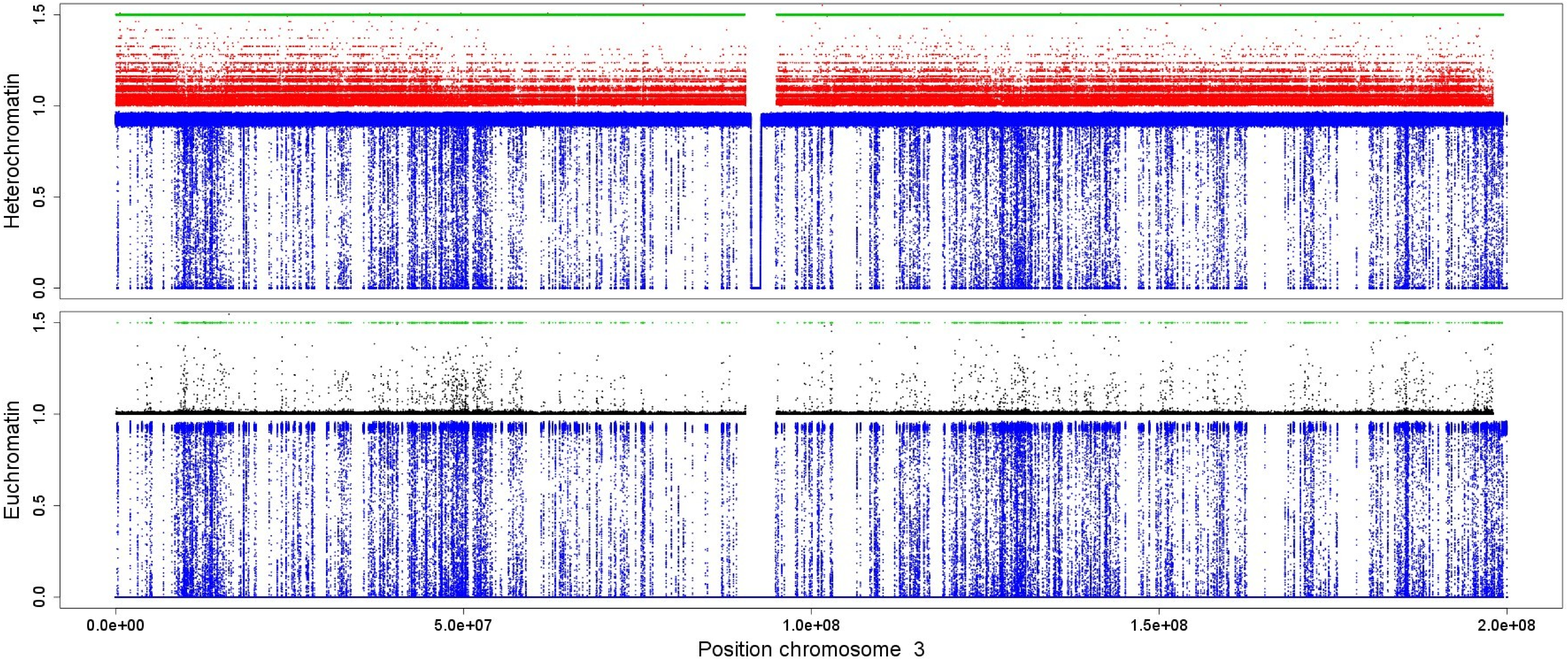}
\includegraphics[width=0.8\textwidth]{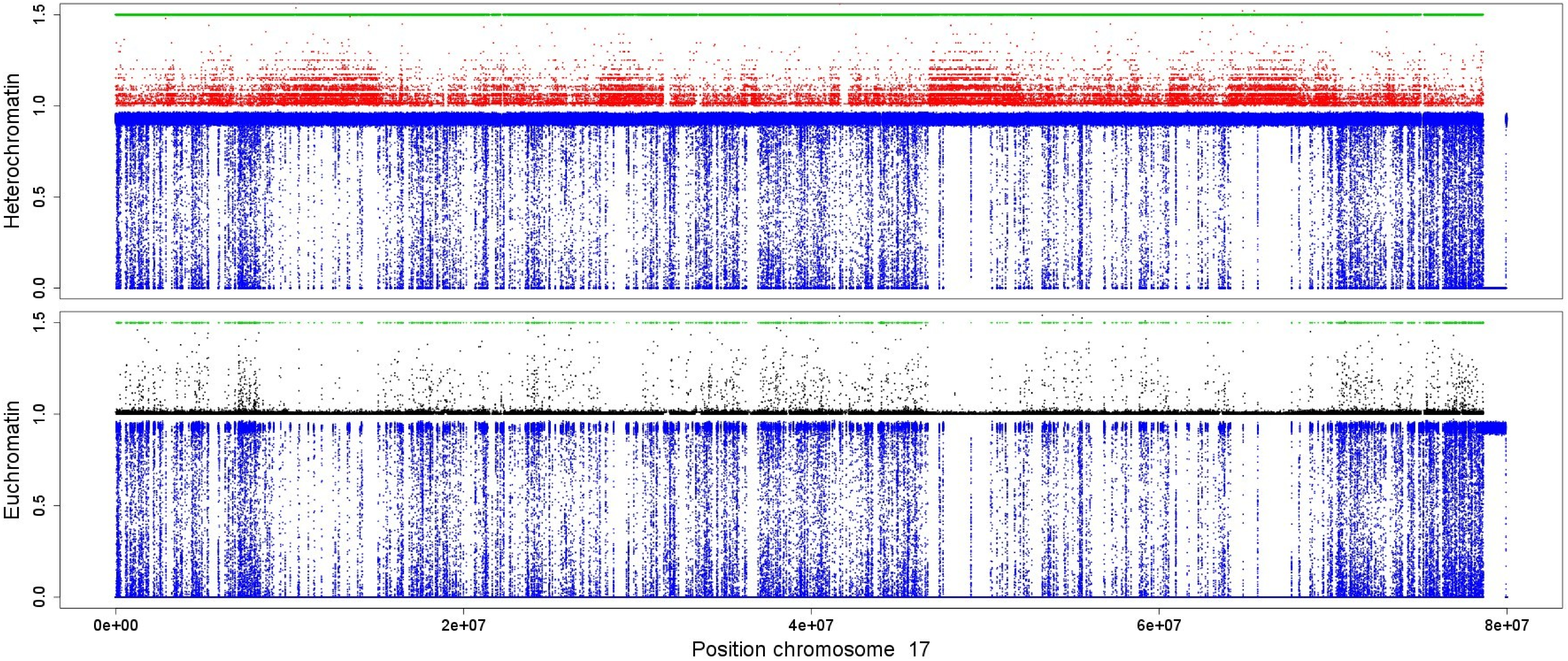}
\caption{Comparing the simulation results for chromosomes 1-3 and 17 to the HeLa cells data set. Parameters are the same as in
Fig.~5.}
\label{fig:CompRealHChr2-17}
\end{figure}
\begin{figure}
\centering
\includegraphics[width=0.9\textwidth]{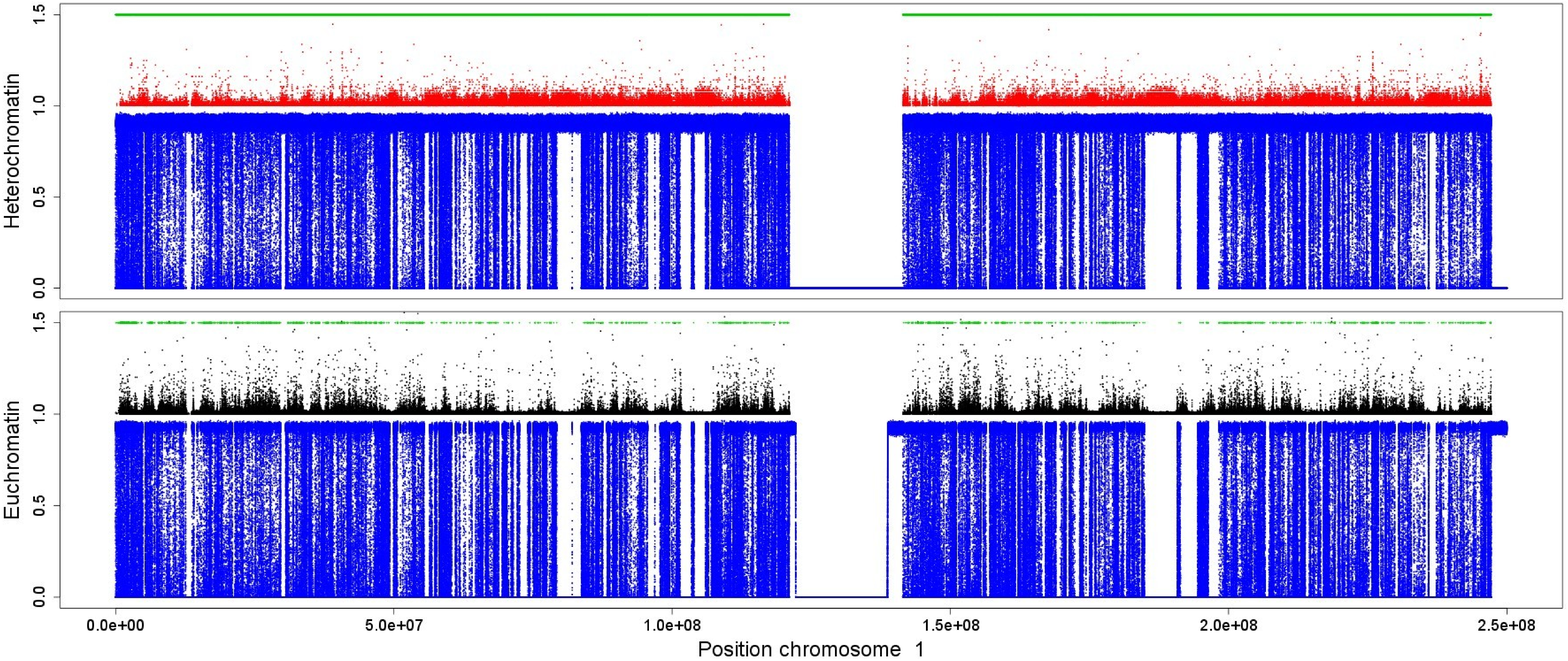}
\caption{Comparing the simulation results for chromosome 3 to the CD4+ ChIP-seq data set. Parameters are the same as in Fig.~5 
except of a smaller propagation rate for heterochromatin, $p_{s,1}=0.08$ leading to an aberrant state of chromatin domain distribution.}
\label{fig:CompRealChr1Aberrant}
\end{figure}
\begin{figure*}
\centering
\psfrag{avlabel}[1][1][4]{ $\boldsymbol{<n_{m}>}$}
\psfrag{fllabel}[1][1][2]{$\boldsymbol{<n^2_{m}>-<n_{m}>^2}$}
\includegraphics[angle=270,width=0.4\textwidth]{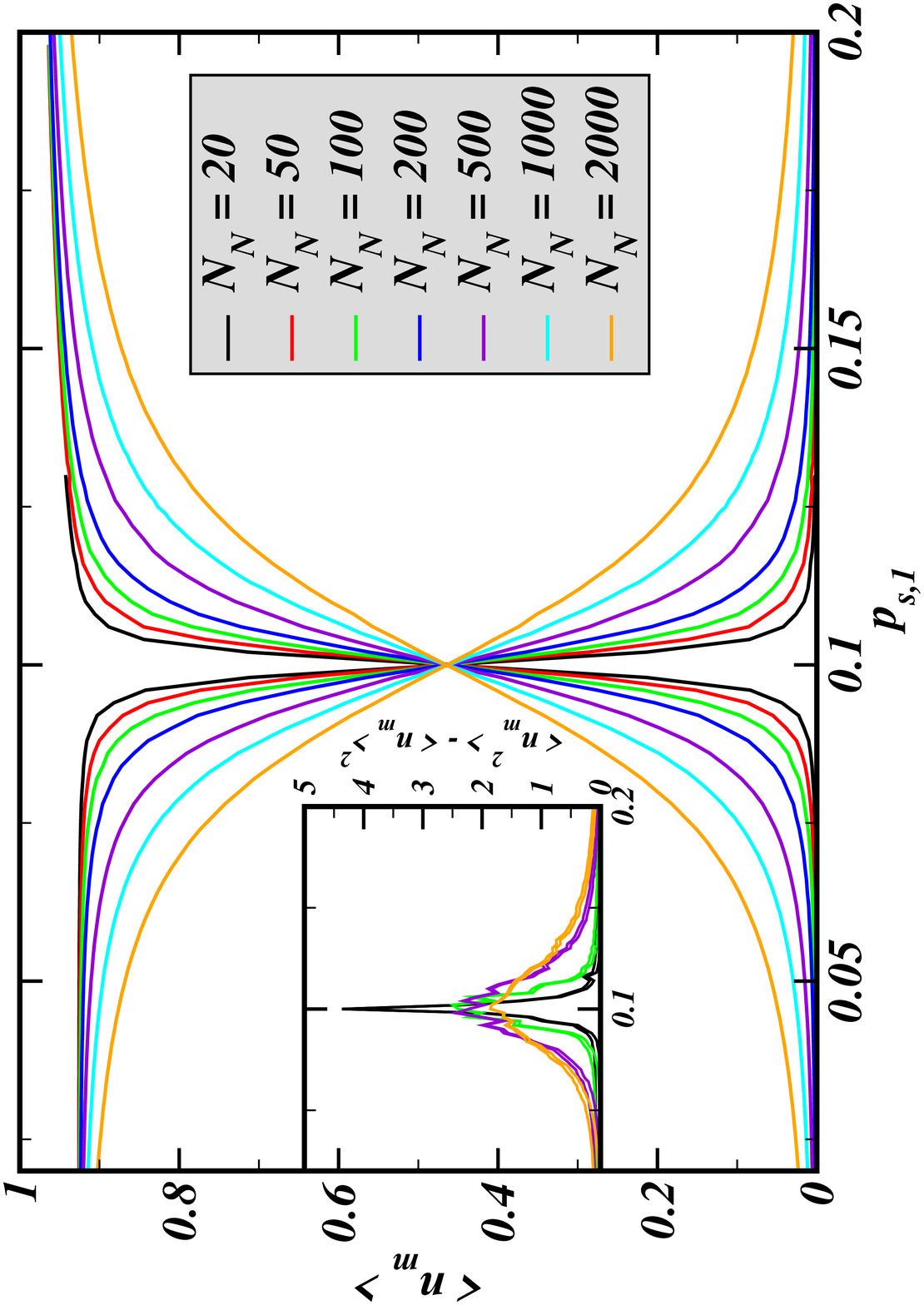}
\hspace{1cm}
\psfrag{avlabel}[1][1][4]{ $\boldsymbol{<n_{m}>}$}
\psfrag{fllabel}[1][1][2]{$\boldsymbol{<n^2_{m}>-<n_{m}>^2}$}
\includegraphics[angle=270,width=0.4\textwidth]{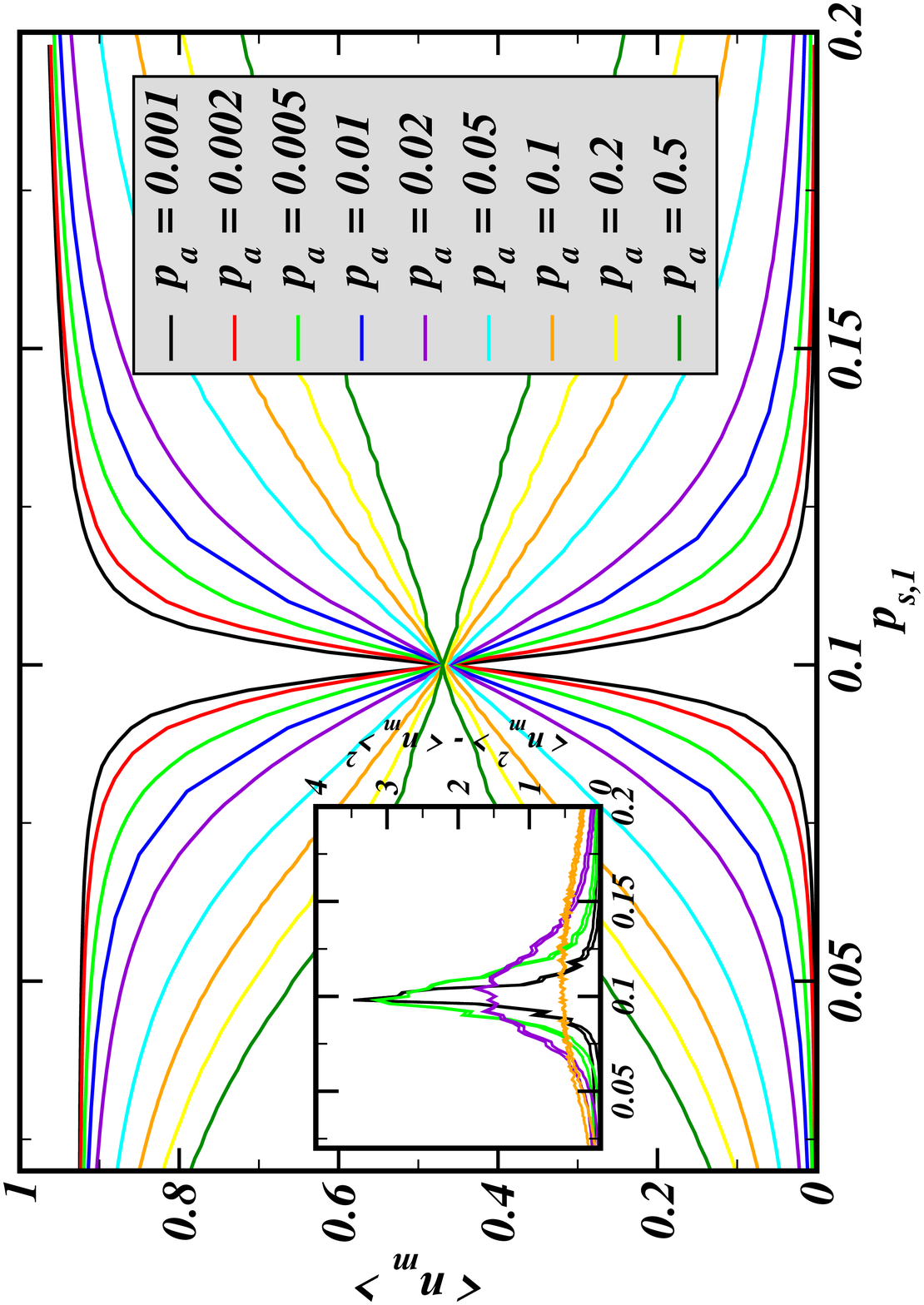}
\caption{General model behavior. Comparing the average frequency of modifications versus the propagation rate $p_{s,1}$ for different numbers of
nucleation sites $N_N$ (left) and for different values of the association rate $p_a$ (right). The
inner panels exhibit the temporal fluctuations $<n_{m}^2>-<n_m>^2$ in the system for each mark. The system exhibits a behavior similar to a phase
transition when changing the propagation constant leading to a drastic increase of the fluctuations at the transition point at $p_{s,1}=p_{s,2}$. At
this point, the domains actively compete against each other by changing their size and temporally occupying regions that have been previously occupied
by the competing mark. The fluctuations become larger for sharper transitions. The transition becomes smoother for smaller numbers of nucleation sites
and/or for larger nucleation rates. The other parameters were $p_{s,2}=0.1$, $p_{d}=p_{a}=0.01$.}
\label{fig:Nm_diffNN_pa}
\end{figure*}
\begin{figure}
\centering
\psfrag{avlabel}[1][1][4]{$\boldsymbol{<n_{m}>}$}
\psfrag{fllabel}[1][1][2]{$\boldsymbol{<n^2_{m}>-<n_{m}>^2}$}
\includegraphics[angle=270,width=0.4\textwidth]{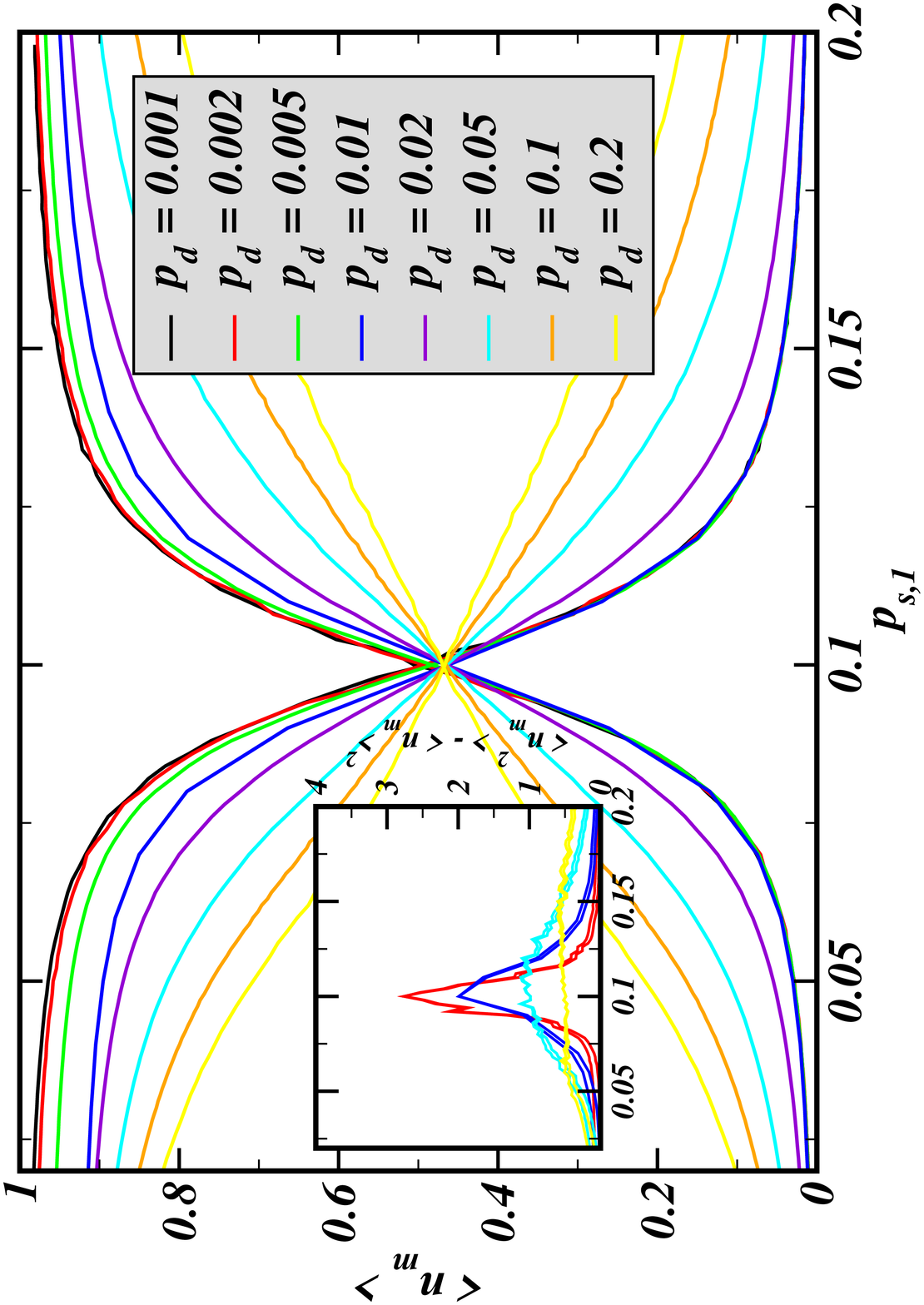}
\caption{General model behavior. Comparing the average frequency of modifications versus the dissociation rate $p_{s,1}$ for different
dissociation rates $p_d$. The other parameters were $p_{s}=0.1,p_{a}=0.01,N_N= 1000$.}
\label{fig:Nm_diffpd}
\end{figure}
\begin{figure*}
\centering
\psfrag{avlabel}[1][1][4]{ $\boldsymbol{<n_{m}>}$}
\includegraphics[angle=270,width=0.31\textwidth]{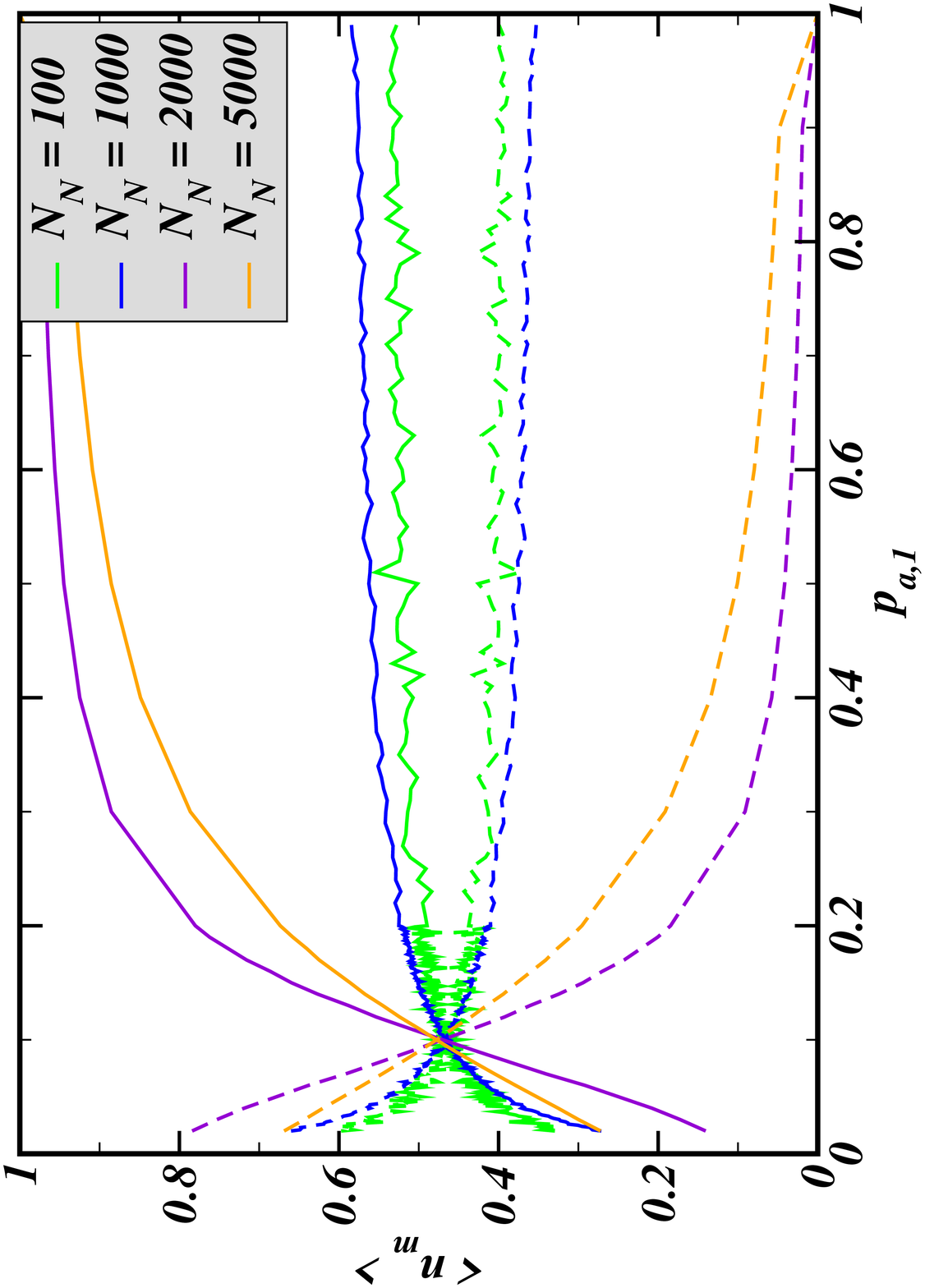}\hspace{0.25cm}
\includegraphics[angle=270,width=0.31\textwidth]{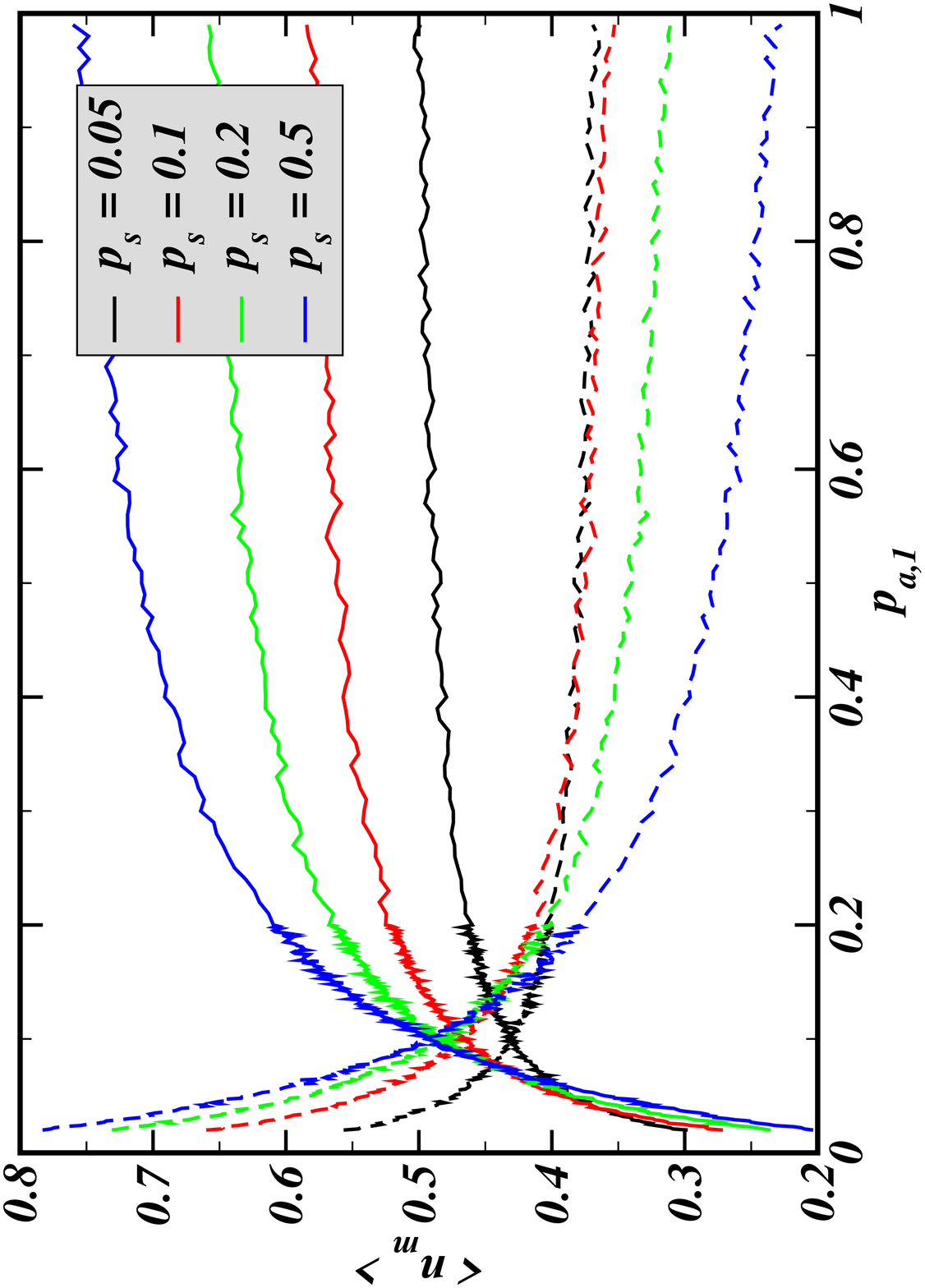}\hspace{0.25cm}
\includegraphics[angle=270,width=0.31\textwidth]{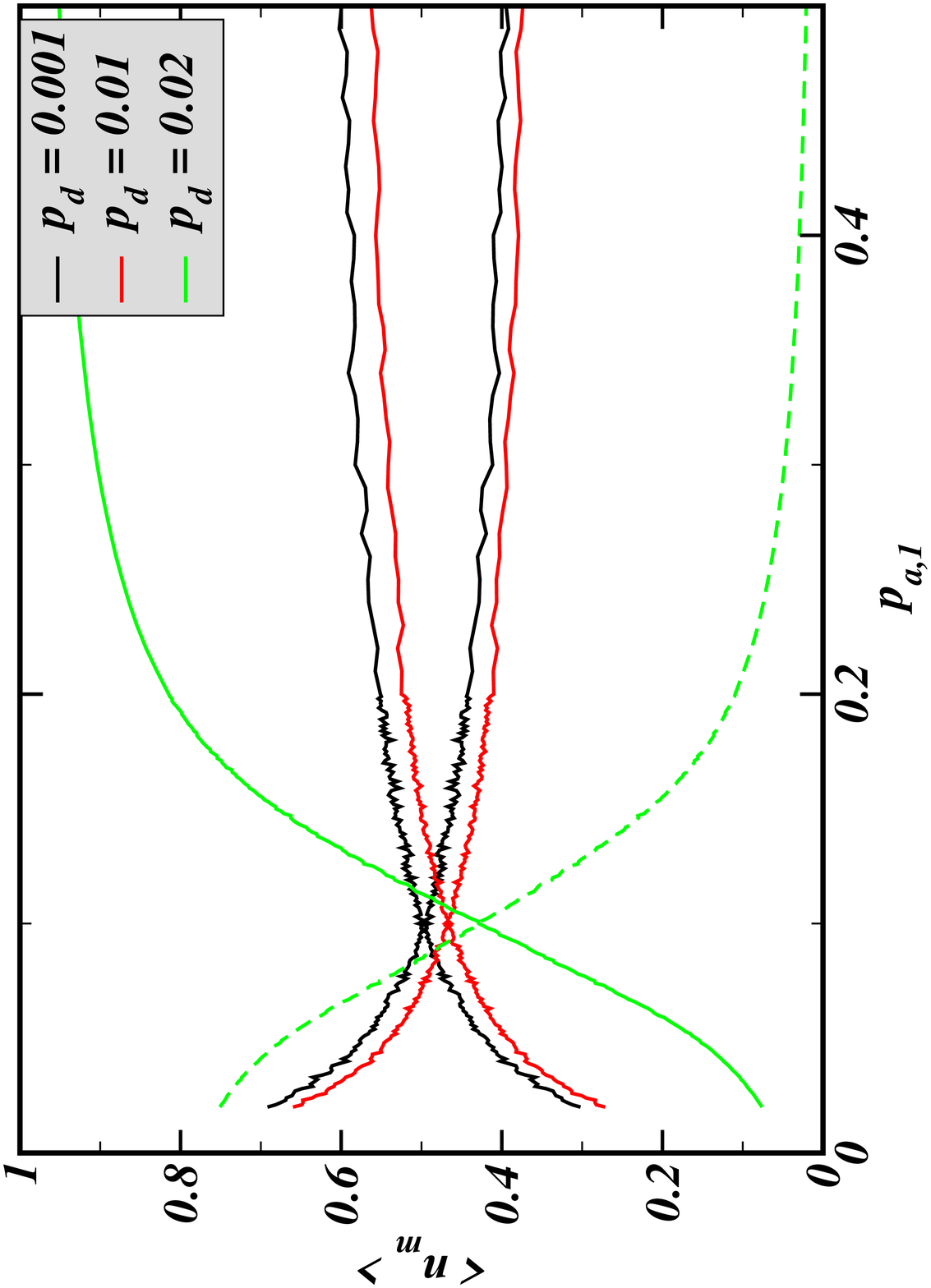}
\caption{Model behavior for different nucleation rates. We compare the average frequency of modifications versus the association rate
$p_{a,1}$ for different numbers of nucleation sites (left), for different values of the propagation rate (center) and for different
values of the deletion rate (right). There is no sensitive
reaction to a change of the association rate.}
\label{fig:Nm_ch_pa}
\end{figure*}

\end{document}